\documentclass[english,prb,10pt,fleqn,floats,secnumarabic,floatfix,twocolumn,superscriptaddress]{revtex4-1}
\usepackage[T1]{fontenc}
\usepackage[latin9]{inputenc}
\usepackage{color}
\usepackage{verbatim}
\usepackage{units}
\usepackage{textcomp}
\usepackage{amsmath}
\usepackage{graphicx}
\usepackage{esint}

\makeatletter

\providecommand{\tabularnewline}{\\}

\usepackage{url}
\usepackage[colorlinks = true,linkcolor = blue,urlcolor = blue,citecolor = blue,anchorcolor = blue]{hyperref}

\usepackage[pagewise,displaymath,mathlines]{lineno}

\makeatother

\usepackage{babel}
\begin{document}

\title{Disorder-sensitive pump-probe measurements on $Nd_{1.83}Ce_{0.17}CuO_{4\pm\delta}$
films}

\author{Adolfo~Avella}

\affiliation{Dipartimento di Fisica ``E.R. Caianiello'', Universit\`a degli
Studi di Salerno, I-84084 Fisciano (SA), Italy}

\affiliation{CNR-SPIN, UoS di Salerno, Via Giovanni Paolo II 132, I-84084 Fisciano
(SA), Italy}

\affiliation{Unit\`a CNISM di Salerno, Universit\`a degli Studi di Salerno,
I-84084 Fisciano (SA), Italy}

\author{Carmela~Buonavolont\`a}

\affiliation{Dipartimento di Fisica ``E. Pancini'', Universit\`a degli Studi
di Napoli \textquotedblleft Federico II\textquotedblright , I-80126
Napoli, Italy}

\author{Anita~Guarino}

\affiliation{Dipartimento di Fisica ``E.R. Caianiello'', Universit\`a degli
Studi di Salerno, I-84084 Fisciano (SA), Italy}

\affiliation{CNR-SPIN, UoS di Salerno, Via Giovanni Paolo II 132, I-84084 Fisciano
(SA), Italy}

\author{Massimo~Valentino}

\affiliation{CNR-SPIN, UoS di Napoli, Complesso di Monte S. Angelo - Via Cinthia,
I-80126 Napoli, Italy}

\author{Antonio~Leo}

\affiliation{Dipartimento di Fisica ``E.R. Caianiello'', Universit\`a degli
Studi di Salerno, I-84084 Fisciano (SA), Italy}

\affiliation{CNR-SPIN, UoS di Salerno, Via Giovanni Paolo II 132, I-84084 Fisciano
(SA), Italy}

\author{Gaia~Grimaldi}

\affiliation{Dipartimento di Fisica ``E.R. Caianiello'', Universit\`a degli
Studi di Salerno, I-84084 Fisciano (SA), Italy}

\affiliation{CNR-SPIN, UoS di Salerno, Via Giovanni Paolo II 132, I-84084 Fisciano
(SA), Italy}

\author{Corrado~de~Lisio}

\affiliation{Dipartimento di Fisica``E. Pancini'', Universit\`a degli Studi
di Napoli \textquotedblleft Federico II\textquotedblright , I-80126
Napoli, Italy}

\affiliation{CNR-SPIN, UoS di Napoli, Complesso di Monte S. Angelo - Via Cinthia,
I-80126 Napoli, Italy}

\author{Angela~Nigro}

\affiliation{Dipartimento di Fisica ``E.R. Caianiello'', Universit\`a degli
Studi di Salerno, I-84084 Fisciano (SA), Italy}

\affiliation{CNR-SPIN, UoS di Salerno, Via Giovanni Paolo II 132, I-84084 Fisciano
(SA), Italy}

\author{Giampiero~Pepe}

\affiliation{Dipartimento di Fisica``E. Pancini'', Universit\`a degli Studi
di Napoli \textquotedblleft Federico II\textquotedblright , I-80126
Napoli, Italy}

\affiliation{CNR-SPIN, UoS di Napoli, Complesso di Monte S. Angelo - Via Cinthia,
I-80126 Napoli, Italy}
\begin{abstract}
We find an unambiguous relationship between disorder-driven features
in the temperature dependence of the resistance and the behavior,
as functions of the temperature, of the parameters necessary to describe
some of the relaxation processes in the photoinduced differential
time-resolved reflectivity of three samples of $Nd_{1.83}Ce_{0.17}CuO_{4\pm\delta}$.
The latter, sharing the same $Ce$ content, have been fabricated and
annealed ad-hoc in order to differ only for the degree of disorder,
mainly related to oxygen content and location, and, consequently,
for the temperature dependence of the resistance: two of them present
a minimum in the resistance and behave as a superconductor and as
a metal, respectively, the third behaves as an insulator. The systematic
coherence between the resistance and the relaxation parameters behaviors
in temperature for all three samples is absolutely remarkable and
shows that pump-probe measurements can be extremely sensitive to disorder
as it drives the emergence of new excitations and of the related relaxation
channels as in this paradigmatic case.
\end{abstract}

\date{\today}

\maketitle

\section{Introduction}

One of the most puzzling peculiarities of superconductivity in electron-doped
cuprates is its relationship to disorder and, in particular, to oxygen
disorder \cite{Armitage_10,Fournier_15}. Superconductivity occurs
in a narrow range of cerium doping and requires an additional oxygen
reduction treatment at the end of the fabrication process \cite{Onose_99,Naito_02,Krockenberger_08,Guarino_13}.
The amount of oxygen removed during this ultimate annealing is very
small \cite{Hirsch_95}, while the treatment has so strong influence
on the electrical and magnetic properties of the samples \cite{Fournier_00,Sekitani_03,Dagan_05,Gauthier_07},
which change from antiferromagnetic insulators to superconductors
\cite{Richard_05,Horio_16}, to suppose that also a redistribution
among all possible oxygen sites could occur as well as some other
structural changes \cite{Riou_04,Kang_07,Guarino_12}. The true effects
of this thermal treatment and its connection to superconductivity
have been the object of intensive experimental studies, but they are
still quite far from being understood and constitute one of the most
intriguing challenges in the field \cite{Guarino_12,Krockenberger_13}.

Different microscopic mechanisms have been conjectured. On one hand,
if oxygen vacancies are created in the $CuO_{2}$ planes and/or in
the charge-reservoir layers, the AF correlation length will decrease\textcolor{red}{{}
}(antiferromagnetic order suppression with respect to as-grown samples
fabricated in oxygen rich atmosphere) as confirmed by angle-resolved
photoemission spectroscopy on differently oxygenated single crystals
\cite{Song_12}. On the other hand, if non-stoichiometric oxygens
at the apical positions are removed, the disorder will be reduced
\cite{Guarino_12}: in optimally reduced crystals, electron localization
and pair breaking effects are reduced with respect to as-grown and
over-reduced samples \cite{Onose_04}. Moreover, the evidence of a
structural restructuring involving copper ion migration \cite{Kang_07}
and the change of the \textit{c} lattice parameter \cite{Guarino_12}
have been reported among the consequences of the high-temperature
oxygen-reduction treatment inducing superconductivity.

One more ingredient in the puzzle is connected to the extreme conditions
in which the annealing procedure necessary to induce superconductivity
has to be carried out \cite{Kim_93}. Accordingly, such a treatment
inevitably becomes itself one of the main sources of disorder in the
system. Such a disorder (mainly related to oxygen vacancies and non-stoichiometric
occupations) largely affect the spectrum of the excitations in the
system and is held responsible for the localization effects observed
in the in-plane transport properties \cite{Barone_09,Barone_11}.
In particular, it was reported that the temperature $T_{m}$ at which
the in-plane resistivity features a minimum in metallic (low-temperature
reduced) and superconducting (high-temperature reduced) samples can
be decreased by repeated low-temperature annealing procedures in oxygen
deficient atmosphere \cite{Guarino_12}. Such a behavior can be easily
explained in terms of a reduction of the degree of positional disorder
of the in-plane oxygen vacancies at each cycle of such a treatment
with a consequent reduction of their (or of any related scattering
center) de-phasing capabilities. Along this line, the high-temperature
treatments (necessary to induce superconductivity) could be instead
held responsible for the removal of non-stoichiometric oxygens at
the apical positions and for the related change of the lattice parameter
in the \textit{c} direction.

Guided by the idea that defects not only affect the spectrum of excitations
of the pristine system, but they can even induce/foster \emph{new}
excitations in the system in connection to their degree of positional
disorder, we have performed time-resolved reflectivity (TRR) pump-probe
experiments on three films of $Nd_{1.83}Ce_{0.17}CuO_{4\pm\delta}$
with substantially different in-plane transport properties induced
by different fabrication procedures and/or ex-situ thermal treatments
with respect to oxygen content and disorder. Our choice of the TRR
technique was dictated by the well-proven capability of pump-probe
techniques to reveal the presence of various kinds of excitations
in a system identifying their different relaxation dynamics \cite{Liu_93,Okamoto_10,Bonavolonta_13,Bonavolonta_14,Cilento_14,Novelli_14,Li_15,Dal-Conte_15,Madan_15,Hinton_16,Vishik_16,Giannetti_16}.
A \emph{new} excitation manifested as \emph{additional} relaxation
channel proving that it is possible to turn a TRR measurement, although
indirectly, into a disorder-sensitive probe. Then, comparing the features
in temperature of the in-plane resistivity of the three samples to
the features in temperature of the \emph{additional }relaxation processes
parametrization, we have managed to prove our initial hypothesis. 

It is necessary to clarify how we have identified the \emph{new} excitations
and the related \emph{additional} relaxation channels: similar TRR
pump-probe experiments have already been performed on a optimally-reduced
single crystal of the same material with a value of cerium doping
equal to $0.15$ \cite{Hinton_13}. In such measurements, at temperatures
above the superconducting critical one, a single relaxation channel
has been identified and assigned to those excitations responsible
for the so-called \emph{pseudogap} phenomenology. Our measurements
reproduce faithfully such results, but also evidence the existence
of a \emph{second} relaxation channel whose parametrization unequivocally
shows features in temperature that can be strictly related to those
reported by the in-plane resistivity of our samples.

Given the main objective to identify \emph{new} excitations and \emph{additional}
relaxation channels to be strictly and unambiguously connected to
oxygen disorder, we have chosen the system to study and the conditions
under which performing this study so to have a well defined reference
and, consequently, to be in the best possible conditions to appreciate
similarities and differences. This is the obvious reason why we have
not only chosen a system where TRR pump-probe experiments have already
been performed \cite{Hinton_13}, without showing any additional channel
with respect to the \emph{canonical} ones (pseudogap and superconductivity),
but we also used the same central wavelength ($\unit[795]{nm}$) and
working pump fluence ($\unit[2]{\mu J/cm^{2}}$). This allowed us
first to benchmark our results by reproducing those already obtained
in Ref.~\onlinecite{Hinton_13} as regards both the pseudogap and
the superconducting channels (the presence of the latter is a trademark
of quality in itself for such systems). Second, in these conditions,
if any different behavior among our films and the single-crystal sample
used in Ref.~\onlinecite{Hinton_13} would come out from our measurements,
it can only be assigned to the main difference between them: the disorder
caused by the different oxygen content and location induced in the
fabrication process.

\section{Theoretical background}

If a system sustains a collective excitation of any kind, the Raman
interaction $H_{\mathrm{R}}$ describes the coupling of an applied
electric field $\mathbf{E}$ (e.g., the one carried by the laser pump
pulse) to such a collective mode \cite{Merlin_97}: $H_{\mathrm{R}}\left(t\right)=\frac{1}{2}\left(\mathbf{E}\left(t\right)\cdot\frac{\partial\boldsymbol{\varepsilon}}{\partial a}\cdot\mathbf{E}\left(t\right)\right)\delta\hat{a}$,
where $a=\left\langle \hat{a}\right\rangle $ is the mean value of
the related \emph{ladder} operator $\hat{a}$, $\delta\hat{a}=\hat{a}-a_{0}$
is the \emph{displacement} operator, $a_{0}=\left\langle \hat{a}\right\rangle _{0}$
is the mean value before the application of the field and $\boldsymbol{\varepsilon}$
is the dielectric tensor of the system. In principle, the dielectric
tensor $\boldsymbol{\varepsilon}$ depends on all collective excitations
of the system and when one of them gets \emph{excited} (i.e., when
$\delta a=\left\langle \delta\hat{a}\right\rangle =a-a_{0}\neq0$)
also $\boldsymbol{\varepsilon}$ gets modified: $\delta\boldsymbol{\varepsilon}=\frac{\partial\boldsymbol{\varepsilon}}{\partial a}\delta a$.
Then, the reflectivity $R$ measured by a second time-delayed pulse
(e.g., the laser probe pulse) is different from the reflectivity measured
at equilibrium (i.e., before the application of the pump pulse) $R_{0}$:
$\Delta R=R-R_{0}\propto\hat{i}\cdot\delta\boldsymbol{\varepsilon}\cdot\hat{r}$,
where $\hat{i}$ and $\hat{r}$ are the incident and reflected directions.
This mechanism is known as impulsive stimulated Raman scattering (ISRS)
\cite{Yan_85} and turns very efficiently TRR measurements into an
invaluable tool to discover collective excitations: $\Delta R\propto\delta a$.
Unfortunately, it is not possible to identify unambiguously the nature
of the mode (spin, charge, orbital, \dots ) in this way, but it is
possible to study the properties of the excitations by analyzing the
features of the corresponding relaxation processes (lifetime of the
mode, type and strength of its damping, \dots ). In fact, within the
linear response regime (i.e., when $\delta a\left(t\right)\cong\int dt'\chi_{a}\left(t-t'\right)\mathsf{F}_{a}\left(t'\right)$
where $\mathsf{F}_{a}\left(t\right)=-\frac{\partial\left\langle H_{\mathrm{R}}\left(t\right)\right\rangle }{\partial a}$)
and for pulses much shorter than the characteristic response time
of the mode (i.e., when $\int dt'\chi_{a}\left(t-t'\right)\mathsf{F}_{a}\left(t'\right)\propto\chi_{a}\left(t\right)$),
the change in reflectivity is proportional to the time-dependent impulsive
response function $\chi_{a}\left(t\right)$ of the excited mode: $\Delta R\left(t\right)\propto\chi_{a}\left(t\right)$.
Accordingly, in presence of one or more critically damped modes (i.e.,
when $\chi_{a}\left(t\right)\propto t\mathrm{e}^{-\gamma_{a}t}$),
the differential transient reflectivity $\tfrac{\Delta R}{R_{0}}\left(T,t\right)$
has the expression: 
\begin{equation}
\tfrac{\Delta R}{R_{0}}\left(T,t\right)=\sum_{i}\Lambda_{i}(T)t\mathrm{e}^{-\nicefrac{t}{\tau_{i}\left(T\right)}}\label{eq:fit}
\end{equation}
Such a behavior, at temperatures above the superconducting critical
one, is exactly what has been found in Ref.~\onlinecite{Hinton_13}
for a single mode that they assigned to the excitations responsible
for the \emph{pseudogap} phenomenology ($\Lambda_{1}$ and $\tau_{1}$).
Here, we search for a second mode of this very same kind whose $\Lambda_{2}$
and $\tau_{2}$ present features that can be related to those of the
in-plane resistivity of our three samples.

\begin{figure}
\noindent \begin{centering}
\includegraphics[width=1\columnwidth]{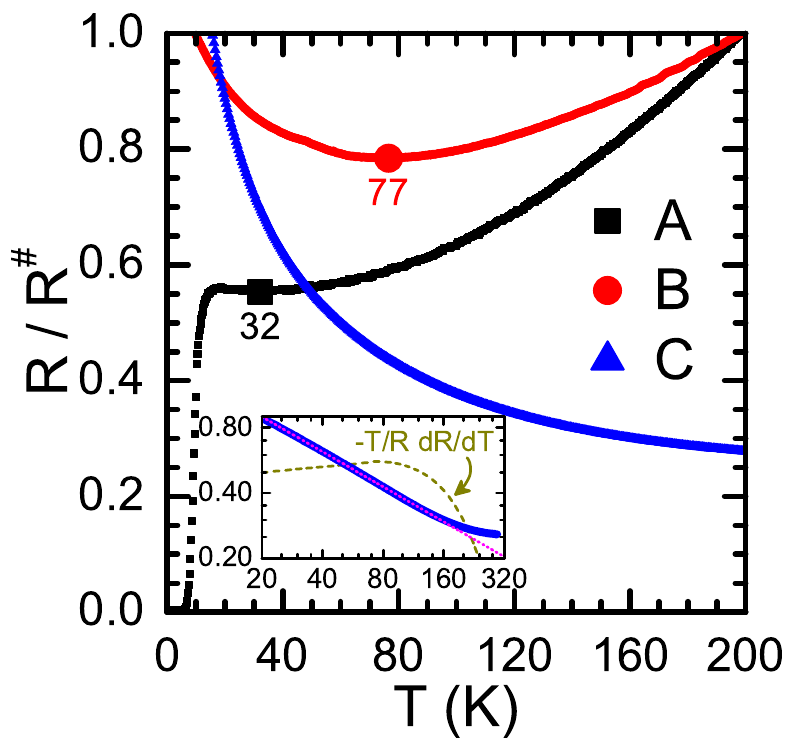}
\par\end{centering}

\caption{Temperature dependence of the resistance of the three samples (scaling
factors $R^{\#}$: $R^{A}=\unit[3.8]{\Omega}$, $R^{B}=\unit[34]{\Omega}$,
and $R^{C}=\unit[300]{\Omega}$). The larger points in A and B mark
the related minimums ($T_{m}^{A}=\unit[77]{K}$, and $T_{m}^{B}=\unit[32]{K}$).
The inset reports the resistance of sample C on a log-log scale together
with the log-log linear fit with slope $-\frac{1}{2}$ (magenta dotted
line) and the opposite of the slope of the log-log tangent to the
curve: $-\frac{T}{R}\frac{dR}{dT}$ (dark yellow dashed line).\label{fig:1}}
\end{figure}

\section{Resistance measurements}

DC sputtering technique was optimized to grow $Nd_{2-x}Ce_{x}CuO_{4\pm\delta}$
films on (100) $SrTiO_{3}$ (STO) substrates by using a single stoichiometric
target as a sputtering source in an on-axis configuration with the
substrate \cite{Guarino_13}. The films were grown in pure argon ($Ar$)
(sample B) or in a mixed atmosphere of $Ar$ and Oxygen ($O_{2}$),
with ratio $O_{2}/Ar>1\%$, at a total pressure of 1.7 mbar (sample
C) and heater temperature of about 850\textdegree C. Part of the sample
B was ex-situ thermally treated in flowing Argon at a temperature
of 900\textdegree C to obtain the sample A, that shows a superconducting
transition at temperature $T{}_{c}\cong\unit[9.8]{K}$ (at 50\% $Rn$)
with $\Delta T_{c}\cong\unit[2]{K}$. The morphology, phase composition,
and purity of the grown samples were inspected by high-resolution
x-ray diffraction and scanning electron microscopy combined with wavelength
dispersive spectroscopy \cite{Guarino_12,Guarino_13}: well oriented
films were obtained without spurious phases and a Cerium content of
$0.17$. The electrical transport properties were investigated by
using the standard four probe technique in a Cryogenic variable-temperature
system. The only difference among the three films, which have been\emph{
}fabricated \emph{ad hoc,} is then the oxygen content and location
that lead to completely different temperature dependences of the in-plane
resistivity {[}see Fig.~\ref{fig:1}{]}: sample A (fabrication in
oxygen-deficient atmosphere, high-temperature oxygen-reduction treatment)
is a \emph{superconductor} ($T{}_{c}\cong\unit[9.8]{K}$), sample
B (fabrication in oxygen-deficient atmosphere, no ex-situ thermal
treatment) is a \emph{metal}, and sample C (fabrication in oxygen-rich
atmosphere, no ex-situ thermal treatment) is an \emph{insulator}.
Samples A and B feature a minimum in the resistivity at temperatures
$T_{m}^{A}=\unit[77]{K}$ and $T_{m}^{B}=\unit[32]{K}$, respectively.
The presence of defects in the system is widely recognized as responsible
for the minimum although the effective type of defects and the scattering
mechanism are still very much debated \cite{Fournier_00,Sekitani_03,Dagan_05,Barone_09}.

\section{Pump-probe measurements}

The pump-probe experimental setup is based on a mode-locked Ti:sapphire
(Ti:Sa) laser, delivering $\unit[100]{fs}$ pulses at a central wavelength
of $\unit[795]{nm}$ (photon energy of $\unit[1.5]{eV}$), at $\unit[82]{MHz}$
repetition rate. The laser beam is split into a p-polarized pump beam,
chopped at $\unit[200]{kHz}$, and a s-polarized probe beam \cite{Bonavolonta_13}.
The pump fluence on the sample ranges from $\unit[2]{\mu J/cm^{2}}$
to $\unit[45.5]{\mu J/cm^{2}}$, while the probe fluence is fixed
at $\unit[2]{\mu J/cm^{2}}$. The sample is located into an optical
cryostat, equipped with a cold finger and using a temperature controlled
liquid helium continuous flow. The measurements were performed in
a temperature range between $\unit[300]{K}$ and $\unit[5]{K}$. For
each temperature, the pump-probe signal is averaged over several fast-scan
sweeps.

\begin{figure}[p]
\noindent \begin{centering}
\begin{tabular}{c}
\includegraphics[height=0.3\textheight]{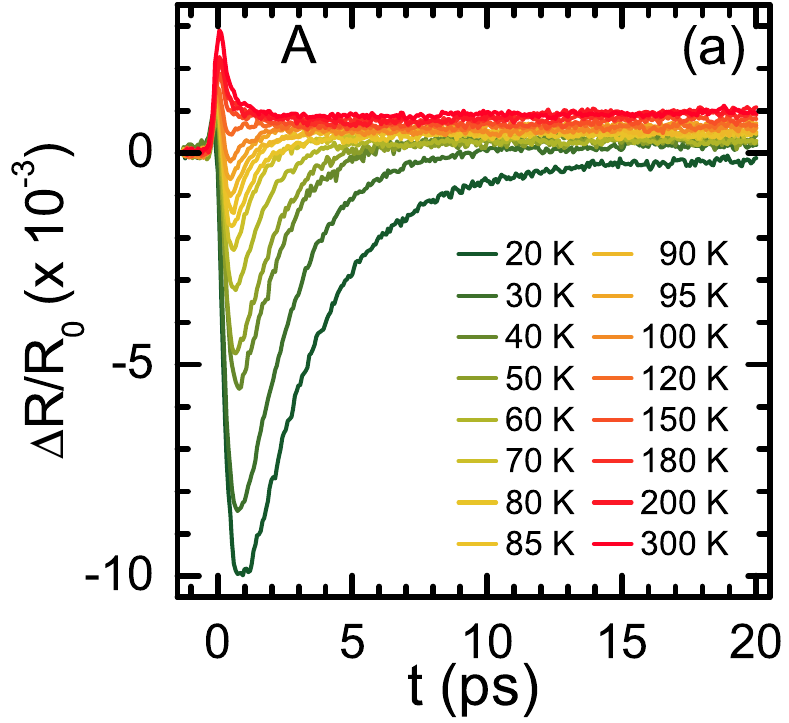}\tabularnewline
\includegraphics[height=0.3\textheight]{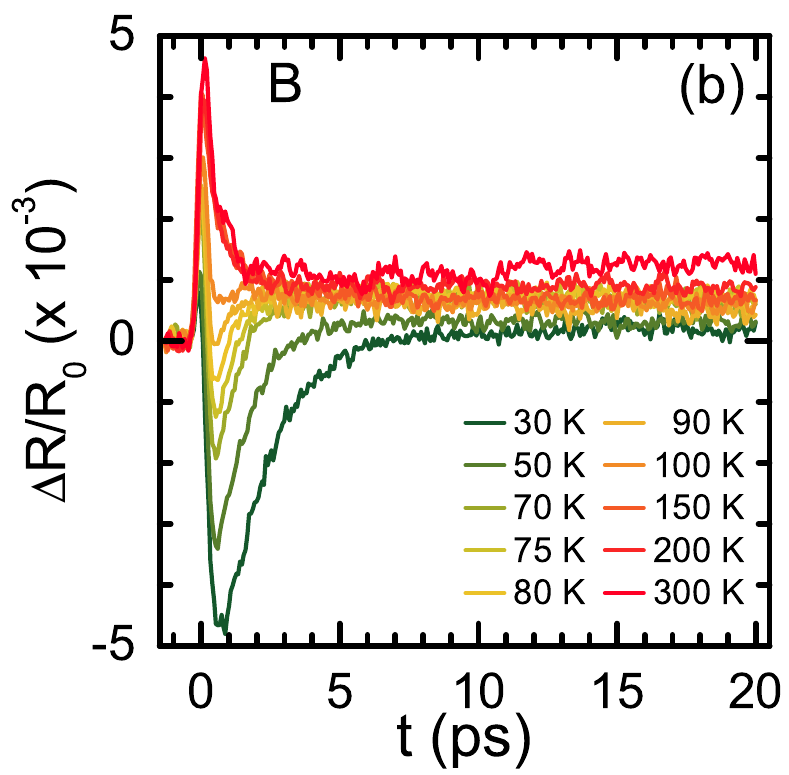}\tabularnewline
\includegraphics[height=0.3\textheight]{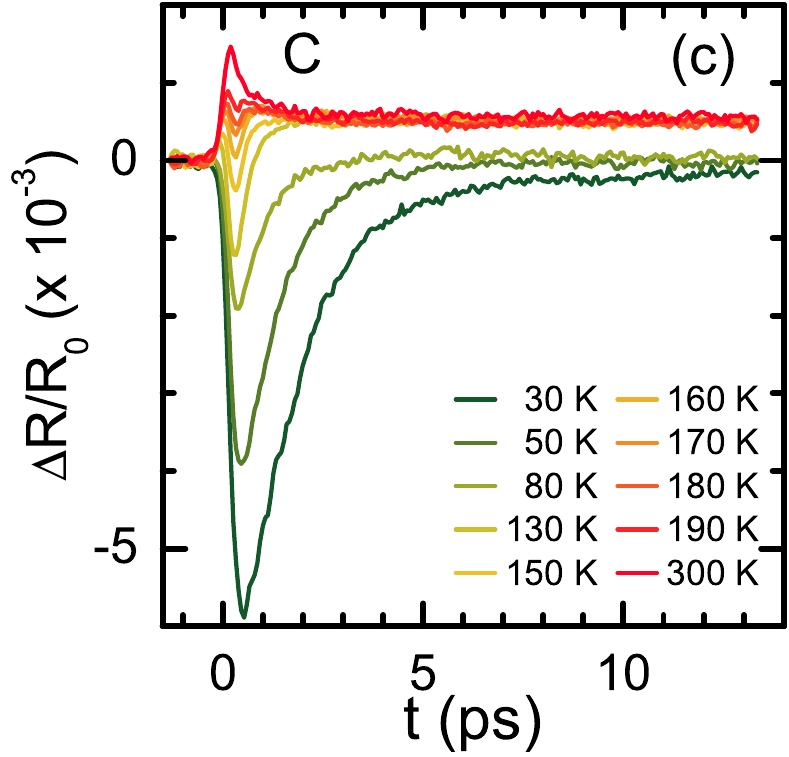}\tabularnewline
\end{tabular}
\par\end{centering}

\caption{Time dependence of the normalized photoinduced transient reflectivity
$\frac{\Delta R}{R_{0}}$ of the three samples {[}(a) A, (b) B, (b)
C{]} at various temperatures and fluence $\Phi=\unit[2]{\mu J/cm^{2}}$.\label{fig:2}}
\end{figure}

\begin{figure}[p]
\noindent \begin{centering}
\begin{tabular}{c}
\includegraphics[height=0.3\textheight]{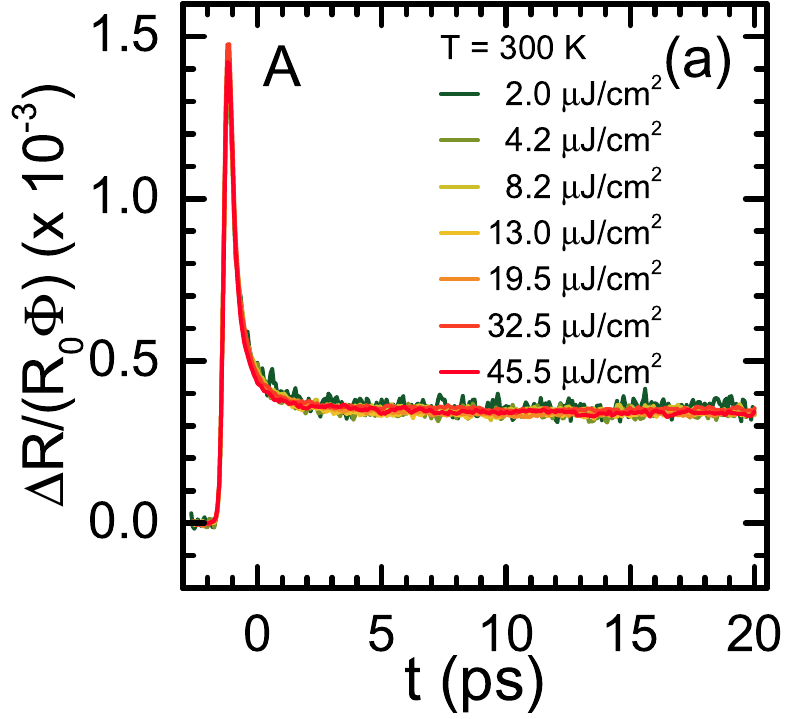}\tabularnewline
\includegraphics[height=0.3\textheight]{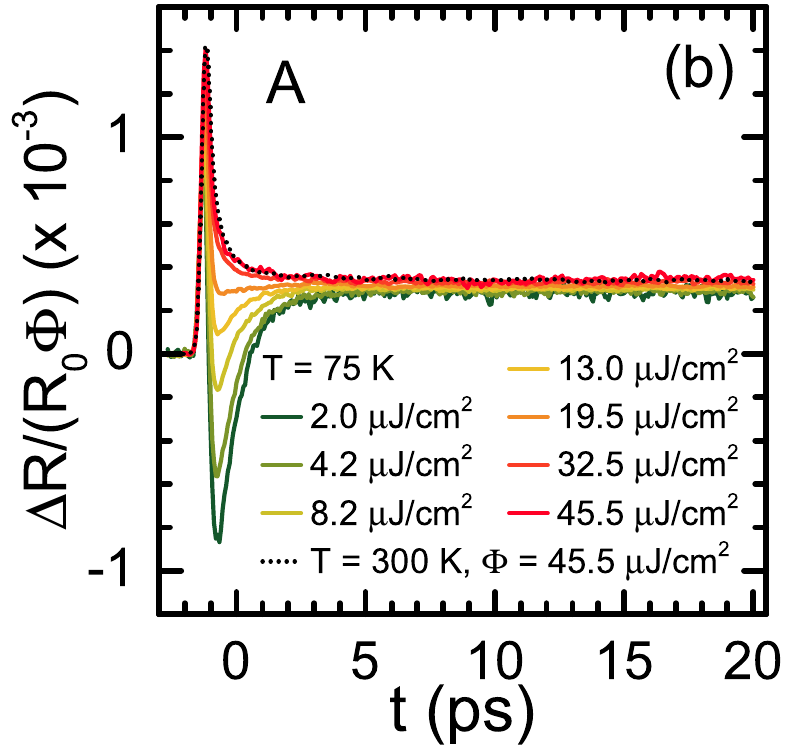}\tabularnewline
\includegraphics[height=0.3\textheight]{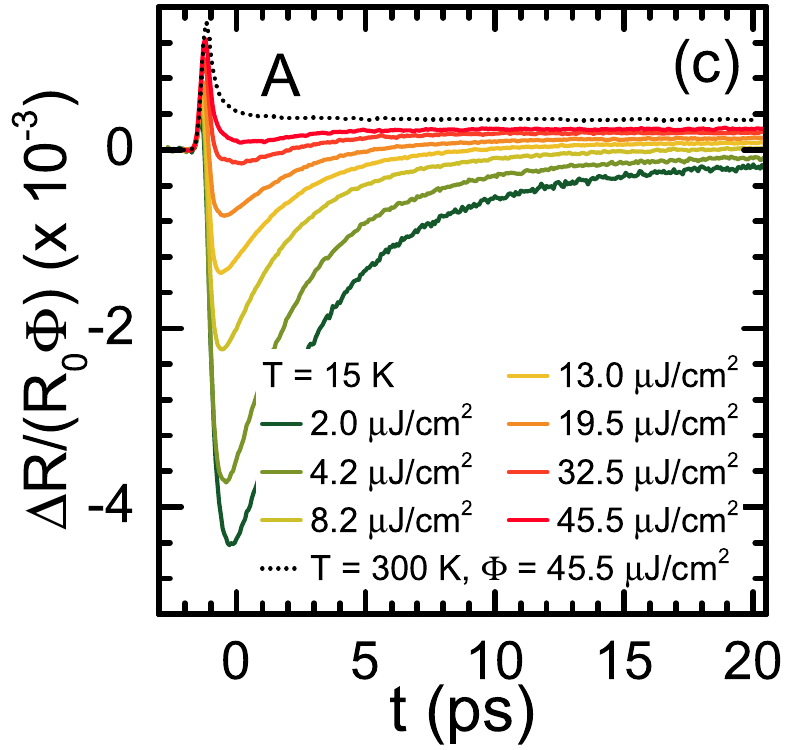}\tabularnewline
\end{tabular}
\par\end{centering}

\caption{Time dependence of the normalized photoinduced transient reflectivity
$\frac{\Delta R}{R_{0}}$ of sample A. At $T=\unit[300]{K}$ (a),
$T=\unit[75]{K}$ (b), $T=\unit[15]{K}$ (c) and for various fluences.
In (b) and (c), $\frac{\Delta R}{R_{0}}$ at $T=\unit[300]{K}$ and
$\Phi=\unit[45.5]{\mu J/cm^{2}}$ is reported as reference (black
dotted line).\label{fig:3}}
\end{figure}

\begin{figure}[p]
\noindent \begin{centering}
\begin{tabular}{c}
\includegraphics[height=0.3\textheight]{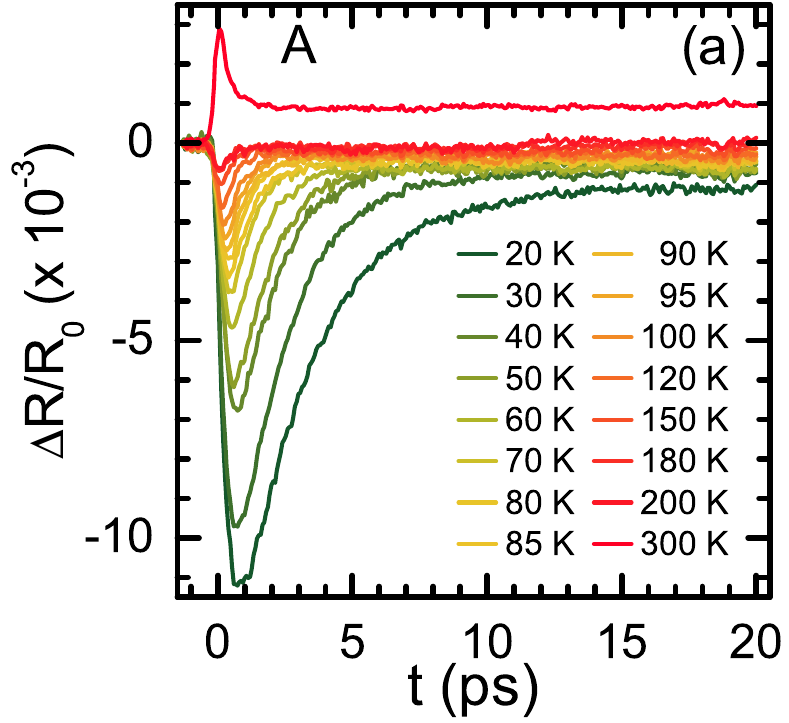}\tabularnewline
\includegraphics[height=0.3\textheight]{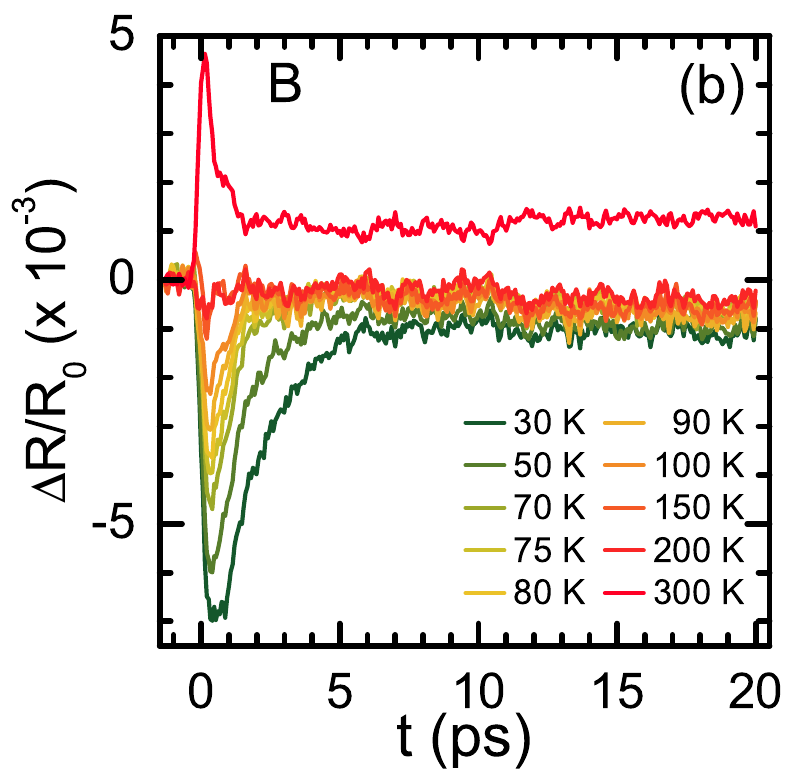}\tabularnewline
\includegraphics[height=0.3\textheight]{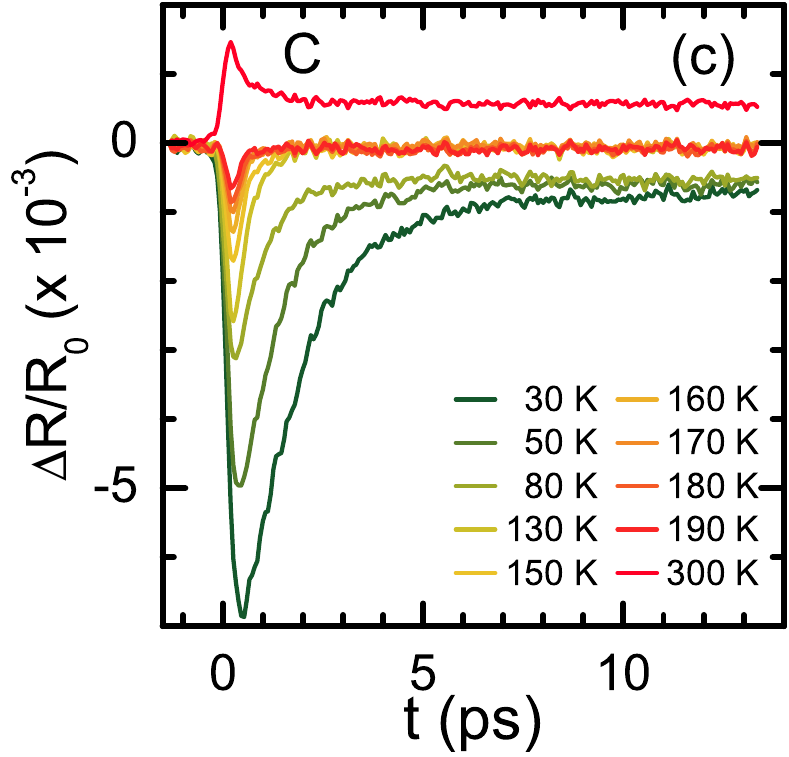}\tabularnewline
\end{tabular}
\par\end{centering}

\caption{Time dependence of the normalized photoinduced transient reflectivity
$\frac{\Delta R}{R_{0}}$ of samples (a) A, (b) B, and (c) C at various
temperatures and $\Phi=\unit[2]{\mu J/cm^{2}}$. The curves for temperatures
lower than $T_{max}=\unit[300]{K}$ have been obtained subtracting
the one at $T_{max}$.\label{fig:4}}
\end{figure}

\begin{figure}[p]
\noindent \begin{centering}
\begin{tabular}{c}
\includegraphics[height=0.3\textheight]{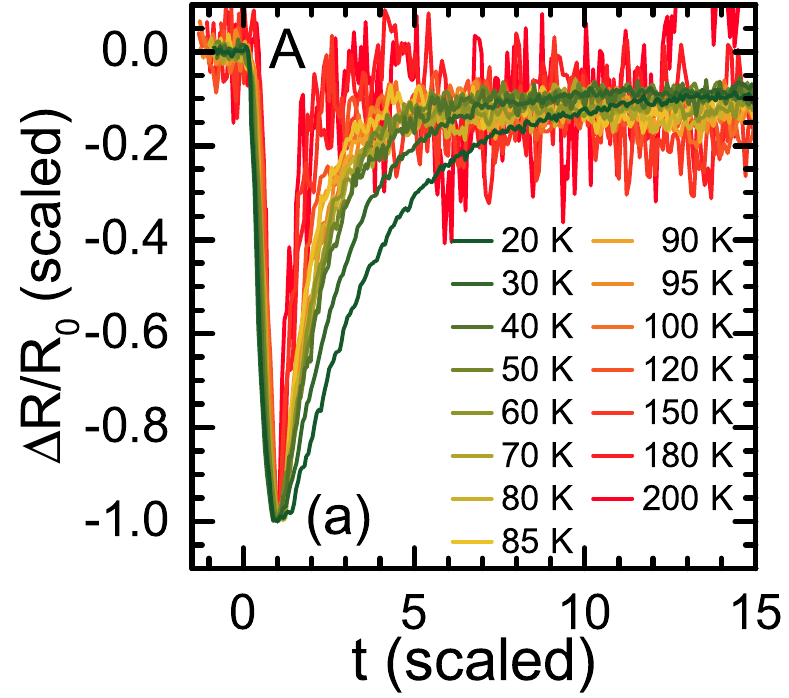}\tabularnewline
\includegraphics[height=0.3\textheight]{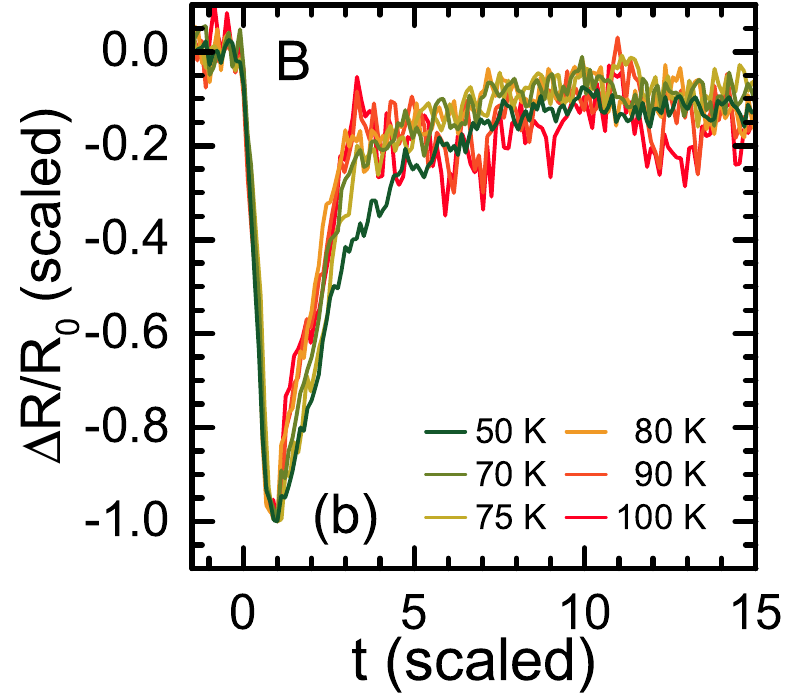}\tabularnewline
\includegraphics[height=0.3\textheight]{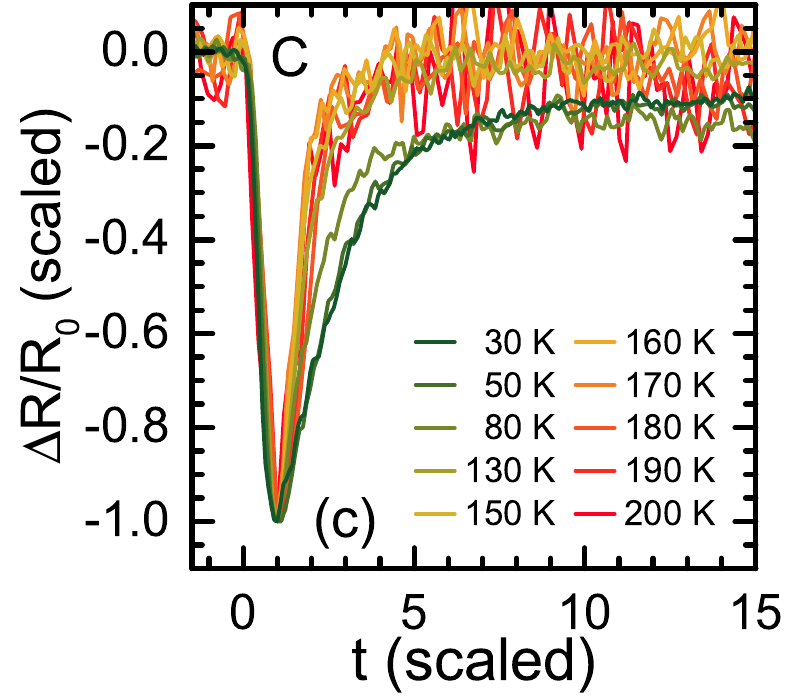}\tabularnewline
\end{tabular}
\par\end{centering}

\caption{Time dependence of the normalized photoinduced transient reflectivity
$\frac{\Delta R}{R_{0}}$ of samples (a) A, (b) B, and (c) C. The
values on both axes have been scaled with respect to the position
of the related (temperature by temperature) minimum.\label{fig:5}}
\end{figure}

\subsection{Thermo-modulation}

The normalized photoinduced transient reflectivity $\tfrac{\Delta R}{R_{0}}$
of the three samples has been measured for various experimental conditions
and is reported in Fig.~\ref{fig:2}. For all three samples, the
most evident feature is the presence of at least two components in
the signal with opposite signs and different temperature dependences
leading to an overall change of the amplitude from positive to negative
on decreasing the temperature. It is worth noting that the positive
component survives also at the lowest temperatures as it can be clearly
seen at the shortest times for all three samples. At the highest measured
temperature $T_{max}=\unit[300]{K}$, the signal is perfectly proportional
to the probe fluence $\Phi$ {[}see Fig.~\ref{fig:3} (a){]} as one
expects if only the thermo-modulation \cite{Brorson_90} component
of the signal would be present. We have exploited this occurrence
to extract the other components of the signal and focus our analysis
on them. In particular, we have subtracted the signal measured at
the temperature $T_{max}$ from the signals measured at all other
temperatures, assuming that the amplitude of the thermo-modulation
component is independent of temperature. This is definitely true at
almost all temperatures in the measured range as clearly shown by
the almost perfect coincidence between the signals in Fig.~\ref{fig:3}
(b) at $T=\unit[75]{K}$ and $T=\unit[300]{K}$ for the maximum value
of the fluence, $\Phi=\unit[45.5]{\mu J/cm^{2}}$, but it will introduce
a systematic error in our analysis at the lowest temperatures {[}see
Fig.~\ref{fig:3} (c){]}. In fact, it is evident that when the temperature
gets smaller and smaller and closer to the range where the superconductivity
manifests, the assumption that the thermo-modulation component of
the signal is independent of temperature is less and less robust and
that higher values of fluence would be necessary to eliminate the
other contributions. In Fig.~\ref{fig:4}, the curves resulting from
the subtraction of the one at the temperature $T_{max}$ are reported
to show the absence of any further component forcing a change of sign
and concavity at the shortest times.

\subsection{Scaling analysis}

Now, a scaling analysis of the data is necessary in order to verify
the assumption that the differential transient reflectivity $\tfrac{\Delta R}{R_{0}}\left(T,t\right)$
can be fitted and analyzed using expression (\ref{eq:fit}). Such
an expression implies that the curves at all temperatures collapse
one on top of the others if the data are scaled, temperature by temperature,
by the coordinates of the related minimum. Actually, the scaling holds
for both the linear decrease (at the shortest times) and the exponential
increase (at longer times) only if just one relaxation channel is
present, as in Ref.~\onlinecite{Hinton_13}, while the presence of
more than one relaxation channel of the same type implied by (\ref{eq:fit})
is signaled by a deviation from the scaling in the exponential increase.
The actual robustness of the scaling at the shortest times {[}see
Fig.~\ref{fig:5}{]} clearly signals that all relaxation channels
are related to critically damped modes and that Eq.~(\ref{eq:fit})
holds. On the other hand, the spreading of the curves at longer times
clearly signals that more than one relaxation channel is present in
the system, very much differently from what found in Ref.~\onlinecite{Hinton_13}.
This can be easily understood in terms of the significative difference
in the relevance of the oxygen disorder between the two sets of samples.

\begin{figure}
\noindent \begin{centering}
\begin{tabular}{c}
\includegraphics[height=0.23\textheight]{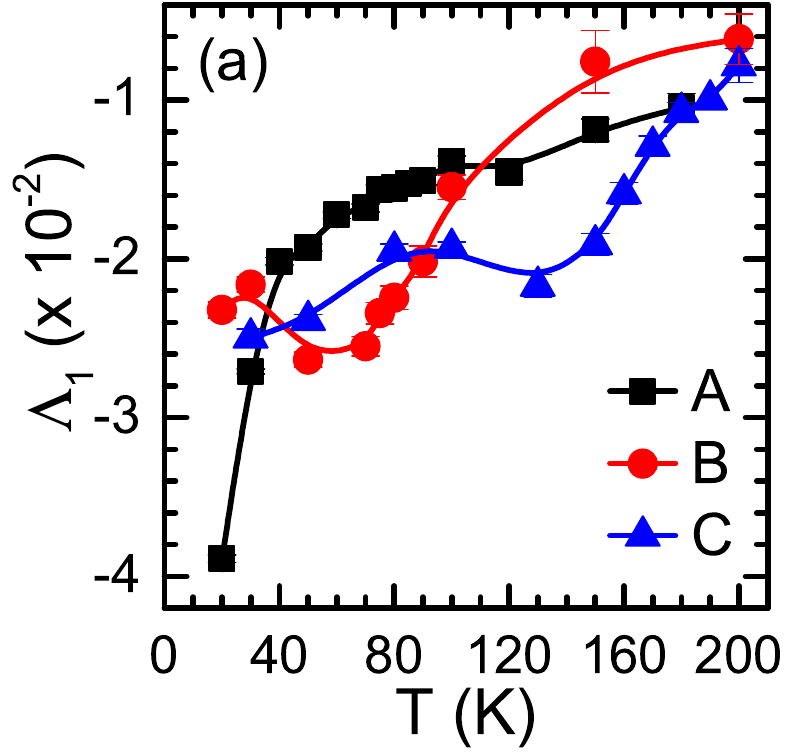}\tabularnewline
\includegraphics[height=0.23\textheight]{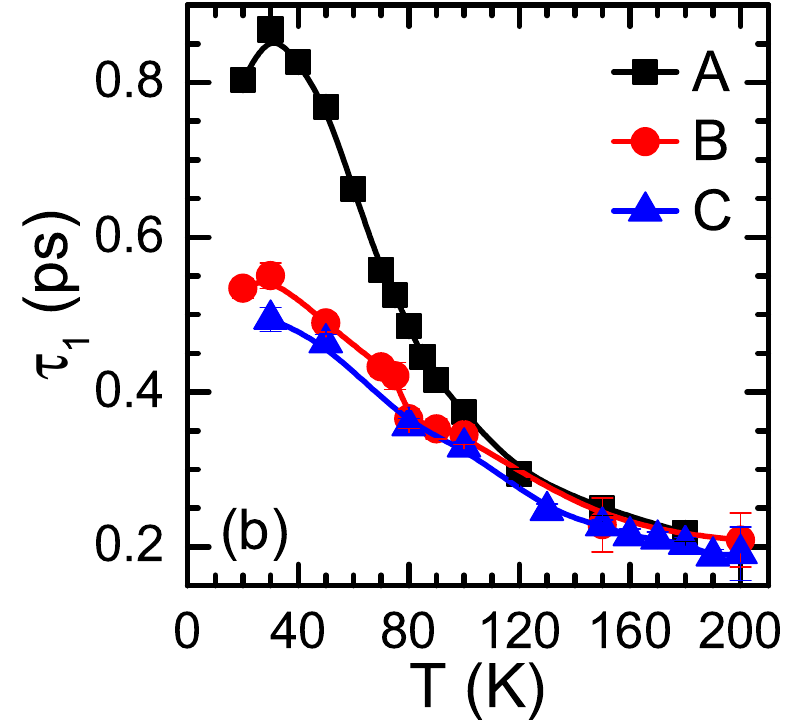}\tabularnewline
\includegraphics[height=0.23\textheight]{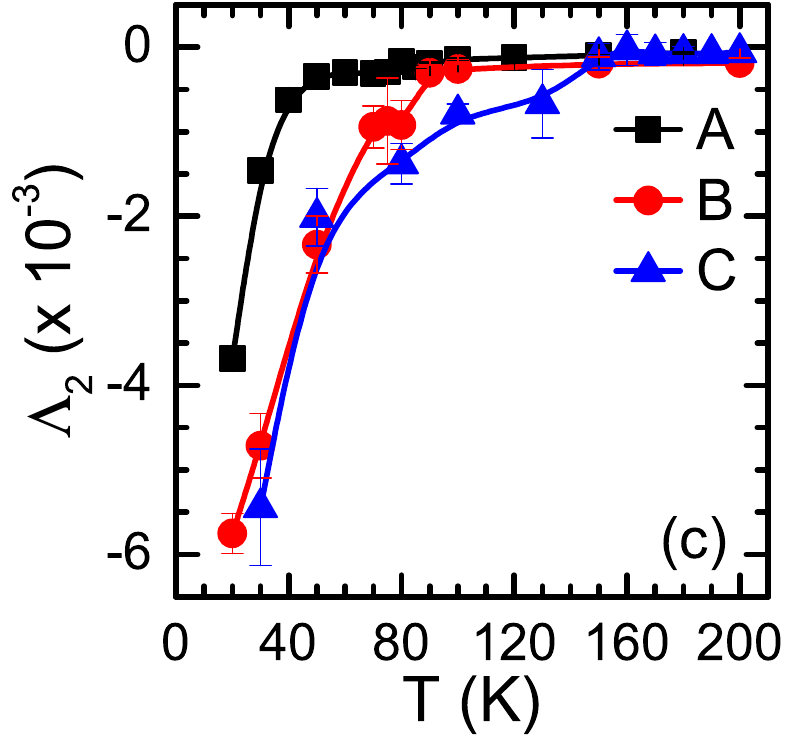}\tabularnewline
\includegraphics[height=0.23\textheight]{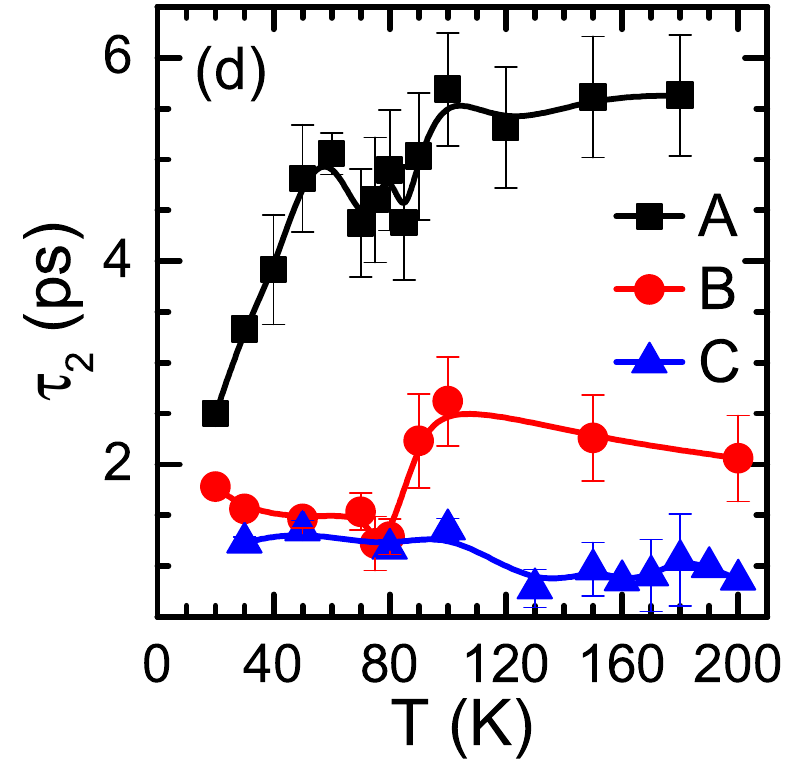}\tabularnewline
\end{tabular}
\par\end{centering}

\caption{Temperature dependence of the main fitting parameters for the three
samples: (a) $\Lambda_{1}$, (b) $\tau_{1}$, (c) $\Lambda_{2}$,
(d) $\tau_{2}$. The lines are guides to the eye.\label{fig:6}}
\end{figure}

\subsection{Relaxation channels analysis}

Given the positive outcome of the scaling analysis for all three samples
(see Fig.~\ref{fig:5}), we can confidently use Eq.~(\ref{eq:fit})
to analyze the data and determine the minimal number of relaxation
channels necessary to fit them. This preliminary analysis shows that
just two channels are active in all three samples and that only four
parameters ($\Lambda_{1},\tau_{1}$ and $\Lambda_{2},\tau_{2}$) are
necessary to fit the data at all analyzed temperatures. In Fig.~\ref{fig:6},
the temperature dependence of these four parameters is reported and
compared among the three different samples. The behavior of $\Lambda_{1}^{A}$
and $\tau_{1}^{A}$ resemble the one found in Ref.~\onlinecite{Hinton_13}.
In particular, we have $\tau_{1}^{A}\left(T\right)\propto\frac{1}{T}$,
which suggests that the excitations responsible for this relaxation
channel are related to a \emph{hidden} order establishing only at
$T=0$. This also suggests that these excitations are the same quasi-particles
lacking global coherence often invoked to explain the opening of a
\emph{pseudogap} in the electronic density of states, reported in
many different experiments, through a dynamic exchange with ordinary
electrons. The first and most remarkable difference with respect to
what found in \cite{Hinton_13} is the change of behavior, on decreasing
temperature, in both $\Lambda_{1}^{A}$ and $\tau_{1}^{A}$ around
the temperature $T_{m}^{A}$, where the resistance features a minimum
{[}see Fig.~\ref{fig:1} (a){]}: $\Lambda_{1}^{A}$ changes suddenly
its slope and $\tau_{1}^{A}$ starts decreasing. The presence of such
a minimum in the resistance is due to disorder and it is evident that
these pump-probe measurements are somehow sensitive to this latter.
In particular, decreasing the temperature, starting right at $T_{m}^{A}$,
the amplitude of the second contribution $\Lambda_{2}^{A}$ becomes
significant {[}see Fig.~\ref{fig:6} (c){]}, marking the occurence
of a second overdamped mode and, correspondingly, of a second relaxation
channel presumably directly connected to disorder. In order to prove
the hypothesis that this second relaxation channel is strictly connected
to disorder, we analyze the behavior of the four fitting parameters
in the other two samples with completely different transport properties.

In sample B, in a region of temperature close to $T_{m}^{B}$, $\Lambda_{1}^{B}$
changes its slope, although less suddenly than $\Lambda_{1}^{A}$
at $T_{m}^{A}$, and $\tau_{1}^{B}\left(T\right)$ changes functional
dependence on temperature: $\propto\frac{1}{T}$ for temperatures
above $T_{m}^{B}$ and $\propto T$ for temperatures below $T_{m}^{B}$.
What is even more worth noticing is that both $\Lambda_{2}^{B}$ becomes
significant and $\tau_{2}^{B}$ jumps from one to another almost constant
value right at $T_{m}^{B}$. Such a jump is present in $\tau_{2}^{A}$
too and, at the lowest temperatures, $\tau_{2}$ assumes almost the
same value in both samples. In sample C, the insulating behavior dominates
the resistive response and no evident minimum is present. Again, $\Lambda_{2}^{C}$
becomes significant only below a certain temperature, $T_{m}^{C}\approx\unit[150]{K}$,
and then shows a behavior remarkably similar to the one reported for
$\Lambda_{2}^{B}$. $\tau_{2}^{B}$ does not show any noticeable change
at that temperature, except for a very small jump in the opposite
direction with respect to those observed in the other two samples,
and assumes almost the same value reported for the other two samples
at low enough temperatures, signaling a kind of \emph{universality}
in the dynamics of this second excitation. As a matter of fact, it
is just the coincidence in the behavior and range of values of $\Lambda_{2}$
and in the values at low temperatures reported for $\tau_{2}$ across
all three samples, although these latter are so different in the resistive
response, to be very remarkable and suggestive of a common behavior
for the slowest relaxation channel.

If we analyze the behavior of $\Lambda_{1}^{C}$ and $\tau_{1}^{C}$,
we can recognize a change in slope in $\Lambda_{1}^{C}$ around $T_{m}^{C}$,
but the overall behavior of $\Lambda_{1}^{C}$ exactly resembles the
one found for $\Lambda_{1}^{B}$ although shifted quite higher in
temperature. Instead, the behavior and the values reported for $\tau_{1}^{C}$
follows very closely those reported for $\tau_{1}^{B}$. The qualitative
and quantitative coincidence in $A_{1}$ and $\tau_{1}$ across all
three samples can be considered as a strong indication that there
exists a strict relationship between the (antiferromagnetic) order
leading to the insulating behavior in sample C and both the \emph{pseudogap}
phenomenon and the fastest relaxation channel in all three samples.

Once the behavior of $\Lambda_{2}^{C}$ and of $\tau_{2}^{C}$ showed
the existence of a kind of critical temperature, $T_{m}^{C}$, also
for sample C, we have searched signatures of it also in the behavior
of the resistance as a function of the temperature. Plotting the resistance
of sample C on a log-log scale {[}see inset in Fig.~\ref{fig:1}(a){]}
and the slope of the log-log tangent to the curve, $-\frac{T}{R}\frac{dR}{dT}$,
in proximity of the temperature $T_{m}^{C}$ found analyzing the optical
data, there appears a change of slope in the resistivity from an almost
constant $-\frac{1}{2}$ at low temperatures to temperature dependent
smaller negative values at higher temperatures. It is worth reminding
that $-\frac{1}{2}$ is what is predicted for an interacting disordered
3D \emph{semiconductor} by weak localization \cite{Lee_85}. This
is the most remarkable confirmation we could seek of the capability
of pump-probe measurements to identify excitations relevant to transport
and of the extreme sensitivity of such measurements.

Starting from a picture where the perfect antiferromagnetic in-plane
long-range spin order in the $Cu$-$O$ planes is weakened (down to
become just short-range) by oxygen vacancies, it is possible to conceive
a unified scenario for both types of excitations revealed by the pump-probe
TRR measurements. The first kind of excitations (those described by
$\Lambda_{1}$ and $\tau_{1}$) could be assigned to the in-plane
magnetic fluctuations reminiscent of the antiferromagnetic order in
the $Cu$-$O$ planes and usually related to the \emph{pseudogap}
phenomenon. Instead, the second kind of excitations (those described
by $\Lambda_{2}$ and $\tau_{2}$) could be assigned to magnetic fluctuations
of the out-of-plane components of the $Cu$ spins induced by the \emph{$O$}
vacancies in the plane. In fact, the absence of an $O$ breaks the
exchange link between the two spins residing on the two nearest-neighbor
$Cu$ ions. This allows them to acquire out-of-plane components with
a specific relative order between them and with respect to all other
couples of out-of-plane spin components generated by $O$ vacancies.
This happens within a distance dictated by the residual correlation
length of the underlying in-plane antiferromagnetic order. The temperature
$T_{m}$ signals the predominance of disorder over all other source
of scattering. The fact that $T_{m}$ can be decreased by repeated
annealing procedures in $O$ deficient atmosphere can be then explained
as such annealing procedures allow the vacancies to maximize their
distances minimizing their pairwise Coulomb potential. Then, the vacancies
somehow order (or, at least, decrease their degree of positional disorder)
with an overall reduction of their de-phasing capability and, therefore,
relevance as scattering sources. According to this, the signals measured
for samples obtained following different procedures of fabrication
and annealing with respect to the $O$ content and positional disorder
is different, but coherent as we found.%

\section{Summary}

In summary, by carefully analyzing the resistance and the TRR of three
\emph{ad hoc} fabricated films of $Nd_{1.83}Ce_{0.17}CuO_{4\pm\delta}$
differing only for the degree of disorder (mainly related to oxygen
vacancies and non-stoichiometric occupations), we have established
the existence of an excitation and of the related critically damped
relaxation channel. The parametrization of this latter presents features
in temperature connected one to one to the transport properties of
the three samples, which are dictated by the presence of defects.
This clearly demonstrates that TRR measurements can be extremely sensitive
to disorder when this latter foster the emergence of new excitations
in the system. A possible scenario for the assignment of the excitations
revealed by the TRR measurements has been proposed that connects the
antiferromagnetic in-plane spin order, its \emph{canonical} fluctuations
inducing a \emph{pseudogap}, its weakening through the insertion of
$O$ vacancies and the effect of the positional disorder of these
latter (controlled through well-defined annealing procedures) on the
appearence and position of a minimum in the resistance as a function
of temperature and the excitations of out-of-plane magnetic fluctuations.
We are quite confident that the collected data and the unambiguous
experimental evidence of a connection between the two kind of measurements
(optical spectroscopy and transport), but also the given possible
scenario, will foster further experiments and investigations.

\acknowledgments

A.A. thanks D. Fausti and C. Giannetti for many insightful discussions.
A.G. and A.L. acknowledge financial support from PON Ricerca e Competitivit\`a
2007-2013 under grant agreement PON NAFASSY, PONa3\_00007.


\begin{thebibliography}{39}%
\makeatletter
\providecommand \@ifxundefined [1]{%
 \@ifx{#1\undefined}
}%
\providecommand \@ifnum [1]{%
 \ifnum #1\expandafter \@firstoftwo
 \else \expandafter \@secondoftwo
 \fi
}%
\providecommand \@ifx [1]{%
 \ifx #1\expandafter \@firstoftwo
 \else \expandafter \@secondoftwo
 \fi
}%
\providecommand \natexlab [1]{#1}%
\providecommand \enquote  [1]{``#1''}%
\providecommand \bibnamefont  [1]{#1}%
\providecommand \bibfnamefont [1]{#1}%
\providecommand \citenamefont [1]{#1}%
\providecommand \href@noop [0]{\@secondoftwo}%
\providecommand \href [0]{\begingroup \@sanitize@url \@href}%
\providecommand \@href[1]{\@@startlink{#1}\@@href}%
\providecommand \@@href[1]{\endgroup#1\@@endlink}%
\providecommand \@sanitize@url [0]{\catcode `\\12\catcode `\$12\catcode
  `\&12\catcode `\#12\catcode `\^12\catcode `\_12\catcode `\%12\relax}%
\providecommand \@@startlink[1]{}%
\providecommand \@@endlink[0]{}%
\providecommand \url  [0]{\begingroup\@sanitize@url \@url }%
\providecommand \@url [1]{\endgroup\@href {#1}{\urlprefix }}%
\providecommand \urlprefix  [0]{URL }%
\providecommand \Eprint [0]{\href }%
\providecommand \doibase [0]{http://dx.doi.org/}%
\providecommand \selectlanguage [0]{\@gobble}%
\providecommand \bibinfo  [0]{\@secondoftwo}%
\providecommand \bibfield  [0]{\@secondoftwo}%
\providecommand \translation [1]{[#1]}%
\providecommand \BibitemOpen [0]{}%
\providecommand \bibitemStop [0]{}%
\providecommand \bibitemNoStop [0]{.\EOS\space}%
\providecommand \EOS [0]{\spacefactor3000\relax}%
\providecommand \BibitemShut  [1]{\csname bibitem#1\endcsname}%
\let\auto@bib@innerbib\@empty
\bibitem [{\citenamefont {Armitage}\ \emph {et~al.}(2010)\citenamefont
  {Armitage}, \citenamefont {Fournier},\ and\ \citenamefont
  {Greene}}]{Armitage_10}%
  \BibitemOpen
  \bibfield  {author} {\bibinfo {author} {\bibfnamefont {N.~P.}\ \bibnamefont
  {Armitage}}, \bibinfo {author} {\bibfnamefont {P.}~\bibnamefont {Fournier}},
  \ and\ \bibinfo {author} {\bibfnamefont {R.~L.}\ \bibnamefont {Greene}},\
  }\href@noop {} {\bibfield  {journal} {\bibinfo  {journal} {Rev. Mod. Phys.}\
  }\textbf {\bibinfo {volume} {82}},\ \bibinfo {pages} {2421} (\bibinfo {year}
  {2010})}\BibitemShut {NoStop}%
\bibitem [{\citenamefont {Fournier}(2015)}]{Fournier_15}%
  \BibitemOpen
  \bibfield  {author} {\bibinfo {author} {\bibfnamefont {P.}~\bibnamefont
  {Fournier}},\ }\href@noop {} {\bibfield  {journal} {\bibinfo  {journal}
  {Physica C}\ }\textbf {\bibinfo {volume} {514}},\ \bibinfo {pages} {314}
  (\bibinfo {year} {2015})}\BibitemShut {NoStop}%
\bibitem [{\citenamefont {Onose}\ \emph {et~al.}(1999)\citenamefont {Onose},
  \citenamefont {Taguchi}, \citenamefont {Ishikawa}, \citenamefont {Shinomori},
  \citenamefont {Ishizaka},\ and\ \citenamefont {Tokura}}]{Onose_99}%
  \BibitemOpen
  \bibfield  {author} {\bibinfo {author} {\bibfnamefont {Y.}~\bibnamefont
  {Onose}}, \bibinfo {author} {\bibfnamefont {Y.}~\bibnamefont {Taguchi}},
  \bibinfo {author} {\bibfnamefont {T.}~\bibnamefont {Ishikawa}}, \bibinfo
  {author} {\bibfnamefont {S.}~\bibnamefont {Shinomori}}, \bibinfo {author}
  {\bibfnamefont {K.}~\bibnamefont {Ishizaka}}, \ and\ \bibinfo {author}
  {\bibfnamefont {Y.}~\bibnamefont {Tokura}},\ }\href@noop {} {\bibfield
  {journal} {\bibinfo  {journal} {Phys. Rev. Lett.}\ }\textbf {\bibinfo
  {volume} {82}},\ \bibinfo {pages} {5120} (\bibinfo {year}
  {1999})}\BibitemShut {NoStop}%
\bibitem [{\citenamefont {Naito}\ \emph {et~al.}(2002)\citenamefont {Naito},
  \citenamefont {Tsukada}, \citenamefont {Greibe},\ and\ \citenamefont
  {Sato}}]{Naito_02}%
  \BibitemOpen
  \bibfield  {author} {\bibinfo {author} {\bibfnamefont {M.}~\bibnamefont
  {Naito}}, \bibinfo {author} {\bibfnamefont {A.}~\bibnamefont {Tsukada}},
  \bibinfo {author} {\bibfnamefont {T.}~\bibnamefont {Greibe}}, \ and\ \bibinfo
  {author} {\bibfnamefont {H.}~\bibnamefont {Sato}},\ }\href@noop {} {\enquote
  {\bibinfo {title} {Phase control in la-214 epitaxial thin films},}\ }
  (\bibinfo {year} {2002})\BibitemShut {NoStop}%
\bibitem [{\citenamefont {Krockenberger}\ \emph {et~al.}(2008)\citenamefont
  {Krockenberger}, \citenamefont {Kurian}, \citenamefont {Winkler},
  \citenamefont {Tsukada}, \citenamefont {Naito},\ and\ \citenamefont
  {Alff}}]{Krockenberger_08}%
  \BibitemOpen
  \bibfield  {author} {\bibinfo {author} {\bibfnamefont {Y.}~\bibnamefont
  {Krockenberger}}, \bibinfo {author} {\bibfnamefont {J.}~\bibnamefont
  {Kurian}}, \bibinfo {author} {\bibfnamefont {A.}~\bibnamefont {Winkler}},
  \bibinfo {author} {\bibfnamefont {A.}~\bibnamefont {Tsukada}}, \bibinfo
  {author} {\bibfnamefont {M.}~\bibnamefont {Naito}}, \ and\ \bibinfo {author}
  {\bibfnamefont {L.}~\bibnamefont {Alff}},\ }\href@noop {} {\bibfield
  {journal} {\bibinfo  {journal} {Phys. Rev. B}\ }\textbf {\bibinfo {volume}
  {77}},\ \bibinfo {pages} {060505} (\bibinfo {year} {2008})}\BibitemShut
  {NoStop}%
\bibitem [{\citenamefont {Guarino}\ \emph {et~al.}(2013)\citenamefont
  {Guarino}, \citenamefont {Patimo}, \citenamefont {Vecchione}, \citenamefont
  {{Di Luccio}},\ and\ \citenamefont {Nigro}}]{Guarino_13}%
  \BibitemOpen
  \bibfield  {author} {\bibinfo {author} {\bibfnamefont {A.}~\bibnamefont
  {Guarino}}, \bibinfo {author} {\bibfnamefont {G.}~\bibnamefont {Patimo}},
  \bibinfo {author} {\bibfnamefont {A.}~\bibnamefont {Vecchione}}, \bibinfo
  {author} {\bibfnamefont {T.}~\bibnamefont {{Di Luccio}}}, \ and\ \bibinfo
  {author} {\bibfnamefont {A.}~\bibnamefont {Nigro}},\ }\href@noop {}
  {\bibfield  {journal} {\bibinfo  {journal} {Physica C}\ }\textbf {\bibinfo
  {volume} {495}},\ \bibinfo {pages} {146} (\bibinfo {year}
  {2013})}\BibitemShut {NoStop}%
\bibitem [{\citenamefont {Hirsch}(1995)}]{Hirsch_95}%
  \BibitemOpen
  \bibfield  {author} {\bibinfo {author} {\bibfnamefont {J.}~\bibnamefont
  {Hirsch}},\ }\href@noop {} {\bibfield  {journal} {\bibinfo  {journal}
  {Physica C}\ }\textbf {\bibinfo {volume} {243}},\ \bibinfo {pages} {319}
  (\bibinfo {year} {1995})}\BibitemShut {NoStop}%
\bibitem [{\citenamefont {Fournier}\ \emph {et~al.}(2000)\citenamefont
  {Fournier}, \citenamefont {Higgins}, \citenamefont {Balci}, \citenamefont
  {Maiser}, \citenamefont {Lobb},\ and\ \citenamefont {Greene}}]{Fournier_00}%
  \BibitemOpen
  \bibfield  {author} {\bibinfo {author} {\bibfnamefont {P.}~\bibnamefont
  {Fournier}}, \bibinfo {author} {\bibfnamefont {J.}~\bibnamefont {Higgins}},
  \bibinfo {author} {\bibfnamefont {H.}~\bibnamefont {Balci}}, \bibinfo
  {author} {\bibfnamefont {E.}~\bibnamefont {Maiser}}, \bibinfo {author}
  {\bibfnamefont {C.~J.}\ \bibnamefont {Lobb}}, \ and\ \bibinfo {author}
  {\bibfnamefont {R.~L.}\ \bibnamefont {Greene}},\ }\href@noop {} {\bibfield
  {journal} {\bibinfo  {journal} {Phys. Rev. B}\ }\textbf {\bibinfo {volume}
  {62}},\ \bibinfo {pages} {R11993} (\bibinfo {year} {2000})}\BibitemShut
  {NoStop}%
\bibitem [{\citenamefont {Sekitani}\ \emph {et~al.}(2003)\citenamefont
  {Sekitani}, \citenamefont {Naito},\ and\ \citenamefont
  {Miura}}]{Sekitani_03}%
  \BibitemOpen
  \bibfield  {author} {\bibinfo {author} {\bibfnamefont {T.}~\bibnamefont
  {Sekitani}}, \bibinfo {author} {\bibfnamefont {M.}~\bibnamefont {Naito}}, \
  and\ \bibinfo {author} {\bibfnamefont {N.}~\bibnamefont {Miura}},\
  }\href@noop {} {\bibfield  {journal} {\bibinfo  {journal} {Phys. Rev. B}\
  }\textbf {\bibinfo {volume} {67}},\ \bibinfo {pages} {174503} (\bibinfo
  {year} {2003})}\BibitemShut {NoStop}%
\bibitem [{\citenamefont {Dagan}\ \emph {et~al.}(2005)\citenamefont {Dagan},
  \citenamefont {Barr}, \citenamefont {Fisher}, \citenamefont {Beck},
  \citenamefont {Dhakal}, \citenamefont {Biswas},\ and\ \citenamefont
  {Greene}}]{Dagan_05}%
  \BibitemOpen
  \bibfield  {author} {\bibinfo {author} {\bibfnamefont {Y.}~\bibnamefont
  {Dagan}}, \bibinfo {author} {\bibfnamefont {M.~C.}\ \bibnamefont {Barr}},
  \bibinfo {author} {\bibfnamefont {W.~M.}\ \bibnamefont {Fisher}}, \bibinfo
  {author} {\bibfnamefont {R.}~\bibnamefont {Beck}}, \bibinfo {author}
  {\bibfnamefont {T.}~\bibnamefont {Dhakal}}, \bibinfo {author} {\bibfnamefont
  {A.}~\bibnamefont {Biswas}}, \ and\ \bibinfo {author} {\bibfnamefont {R.~L.}\
  \bibnamefont {Greene}},\ }\href@noop {} {\bibfield  {journal} {\bibinfo
  {journal} {Phys. Rev. Lett.}\ }\textbf {\bibinfo {volume} {94}},\ \bibinfo
  {pages} {057005} (\bibinfo {year} {2005})}\BibitemShut {NoStop}%
\bibitem [{\citenamefont {Gauthier}\ \emph {et~al.}(2007)\citenamefont
  {Gauthier}, \citenamefont {Gagn\'e}, \citenamefont {Renaud}, \citenamefont
  {Gosselin}, \citenamefont {Fournier},\ and\ \citenamefont
  {Richard}}]{Gauthier_07}%
  \BibitemOpen
  \bibfield  {author} {\bibinfo {author} {\bibfnamefont {J.}~\bibnamefont
  {Gauthier}}, \bibinfo {author} {\bibfnamefont {S.}~\bibnamefont {Gagn\'e}},
  \bibinfo {author} {\bibfnamefont {J.}~\bibnamefont {Renaud}}, \bibinfo
  {author} {\bibfnamefont {M.-E.}\ \bibnamefont {Gosselin}}, \bibinfo {author}
  {\bibfnamefont {P.}~\bibnamefont {Fournier}}, \ and\ \bibinfo {author}
  {\bibfnamefont {P.}~\bibnamefont {Richard}},\ }\href@noop {} {\bibfield
  {journal} {\bibinfo  {journal} {Phys. Rev. B}\ }\textbf {\bibinfo {volume}
  {75}},\ \bibinfo {pages} {024424} (\bibinfo {year} {2007})}\BibitemShut
  {NoStop}%
\bibitem [{\citenamefont {Richard}\ \emph {et~al.}(2005)\citenamefont
  {Richard}, \citenamefont {Poirier}, \citenamefont {Jandl},\ and\
  \citenamefont {Fournier}}]{Richard_05}%
  \BibitemOpen
  \bibfield  {author} {\bibinfo {author} {\bibfnamefont {P.}~\bibnamefont
  {Richard}}, \bibinfo {author} {\bibfnamefont {M.}~\bibnamefont {Poirier}},
  \bibinfo {author} {\bibfnamefont {S.}~\bibnamefont {Jandl}}, \ and\ \bibinfo
  {author} {\bibfnamefont {P.}~\bibnamefont {Fournier}},\ }\href@noop {}
  {\bibfield  {journal} {\bibinfo  {journal} {Phys. Rev. B}\ }\textbf {\bibinfo
  {volume} {72}},\ \bibinfo {pages} {184514} (\bibinfo {year}
  {2005})}\BibitemShut {NoStop}%
\bibitem [{\citenamefont {Horio}\ \emph {et~al.}(2016)\citenamefont {Horio},
  \citenamefont {Adachi}, \citenamefont {Mori}, \citenamefont {Takahashi},
  \citenamefont {Yoshida}, \citenamefont {Suzuki}, \citenamefont {Ambolode~II},
  \citenamefont {Okazaki}, \citenamefont {Ono}, \citenamefont {Kumigashira},
  \citenamefont {Anzai}, \citenamefont {Arita}, \citenamefont {Namatame},
  \citenamefont {Taniguchi}, \citenamefont {Ootsuki}, \citenamefont {Sawada},
  \citenamefont {Takahashi}, \citenamefont {Mizokawa}, \citenamefont {Koike},\
  and\ \citenamefont {Fujimori}}]{Horio_16}%
  \BibitemOpen
  \bibfield  {author} {\bibinfo {author} {\bibfnamefont {M.}~\bibnamefont
  {Horio}}, \bibinfo {author} {\bibfnamefont {T.}~\bibnamefont {Adachi}},
  \bibinfo {author} {\bibfnamefont {Y.}~\bibnamefont {Mori}}, \bibinfo {author}
  {\bibfnamefont {A.}~\bibnamefont {Takahashi}}, \bibinfo {author}
  {\bibfnamefont {T.}~\bibnamefont {Yoshida}}, \bibinfo {author} {\bibfnamefont
  {H.}~\bibnamefont {Suzuki}}, \bibinfo {author} {\bibfnamefont {L.~C.~C.}\
  \bibnamefont {Ambolode~II}}, \bibinfo {author} {\bibfnamefont
  {K.}~\bibnamefont {Okazaki}}, \bibinfo {author} {\bibfnamefont
  {K.}~\bibnamefont {Ono}}, \bibinfo {author} {\bibfnamefont {H.}~\bibnamefont
  {Kumigashira}}, \bibinfo {author} {\bibfnamefont {H.}~\bibnamefont {Anzai}},
  \bibinfo {author} {\bibfnamefont {M.}~\bibnamefont {Arita}}, \bibinfo
  {author} {\bibfnamefont {H.}~\bibnamefont {Namatame}}, \bibinfo {author}
  {\bibfnamefont {M.}~\bibnamefont {Taniguchi}}, \bibinfo {author}
  {\bibfnamefont {D.}~\bibnamefont {Ootsuki}}, \bibinfo {author} {\bibfnamefont
  {K.}~\bibnamefont {Sawada}}, \bibinfo {author} {\bibfnamefont
  {M.}~\bibnamefont {Takahashi}}, \bibinfo {author} {\bibfnamefont
  {T.}~\bibnamefont {Mizokawa}}, \bibinfo {author} {\bibfnamefont
  {Y.}~\bibnamefont {Koike}}, \ and\ \bibinfo {author} {\bibfnamefont
  {A.}~\bibnamefont {Fujimori}},\ }\href@noop {} {\bibfield  {journal}
  {\bibinfo  {journal} {Nat Commun}\ }\textbf {\bibinfo {volume} {7}} (\bibinfo
  {year} {2016})}\BibitemShut {NoStop}%
\bibitem [{\citenamefont {Riou}\ \emph {et~al.}(2004)\citenamefont {Riou},
  \citenamefont {Richard}, \citenamefont {Jandl}, \citenamefont {Poirier},
  \citenamefont {Fournier}, \citenamefont {Nekvasil}, \citenamefont {Barilo},\
  and\ \citenamefont {Kurnevich}}]{Riou_04}%
  \BibitemOpen
  \bibfield  {author} {\bibinfo {author} {\bibfnamefont {G.}~\bibnamefont
  {Riou}}, \bibinfo {author} {\bibfnamefont {P.}~\bibnamefont {Richard}},
  \bibinfo {author} {\bibfnamefont {S.}~\bibnamefont {Jandl}}, \bibinfo
  {author} {\bibfnamefont {M.}~\bibnamefont {Poirier}}, \bibinfo {author}
  {\bibfnamefont {P.}~\bibnamefont {Fournier}}, \bibinfo {author}
  {\bibfnamefont {V.}~\bibnamefont {Nekvasil}}, \bibinfo {author}
  {\bibfnamefont {S.~N.}\ \bibnamefont {Barilo}}, \ and\ \bibinfo {author}
  {\bibfnamefont {L.~A.}\ \bibnamefont {Kurnevich}},\ }\href@noop {} {\bibfield
   {journal} {\bibinfo  {journal} {Phys. Rev. B}\ }\textbf {\bibinfo {volume}
  {69}},\ \bibinfo {pages} {024511} (\bibinfo {year} {2004})}\BibitemShut
  {NoStop}%
\bibitem [{\citenamefont {Kang}\ \emph {et~al.}(2007)\citenamefont {Kang},
  \citenamefont {Dai}, \citenamefont {Campbell}, \citenamefont {Chupas},
  \citenamefont {Rosenkranz}, \citenamefont {Lee}, \citenamefont {Huang},
  \citenamefont {Li}, \citenamefont {Komiya},\ and\ \citenamefont
  {Ando}}]{Kang_07}%
  \BibitemOpen
  \bibfield  {author} {\bibinfo {author} {\bibfnamefont {H.~J.}\ \bibnamefont
  {Kang}}, \bibinfo {author} {\bibfnamefont {P.}~\bibnamefont {Dai}}, \bibinfo
  {author} {\bibfnamefont {B.~J.}\ \bibnamefont {Campbell}}, \bibinfo {author}
  {\bibfnamefont {P.~J.}\ \bibnamefont {Chupas}}, \bibinfo {author}
  {\bibfnamefont {S.}~\bibnamefont {Rosenkranz}}, \bibinfo {author}
  {\bibfnamefont {P.~L.}\ \bibnamefont {Lee}}, \bibinfo {author} {\bibfnamefont
  {Q.}~\bibnamefont {Huang}}, \bibinfo {author} {\bibfnamefont
  {S.}~\bibnamefont {Li}}, \bibinfo {author} {\bibfnamefont {S.}~\bibnamefont
  {Komiya}}, \ and\ \bibinfo {author} {\bibfnamefont {Y.}~\bibnamefont
  {Ando}},\ }\href@noop {} {\bibfield  {journal} {\bibinfo  {journal} {Nat.
  Mater.}\ }\textbf {\bibinfo {volume} {6}},\ \bibinfo {pages} {224} (\bibinfo
  {year} {2007})}\BibitemShut {NoStop}%
\bibitem [{\citenamefont {Guarino}\ \emph {et~al.}(2012)\citenamefont
  {Guarino}, \citenamefont {Fittipaldi}, \citenamefont {Romano}, \citenamefont
  {Vecchione},\ and\ \citenamefont {Nigro}}]{Guarino_12}%
  \BibitemOpen
  \bibfield  {author} {\bibinfo {author} {\bibfnamefont {A.}~\bibnamefont
  {Guarino}}, \bibinfo {author} {\bibfnamefont {R.}~\bibnamefont {Fittipaldi}},
  \bibinfo {author} {\bibfnamefont {A.}~\bibnamefont {Romano}}, \bibinfo
  {author} {\bibfnamefont {A.}~\bibnamefont {Vecchione}}, \ and\ \bibinfo
  {author} {\bibfnamefont {A.}~\bibnamefont {Nigro}},\ }\href@noop {}
  {\bibfield  {journal} {\bibinfo  {journal} {Thin Solid Films}\ }\textbf
  {\bibinfo {volume} {524}},\ \bibinfo {pages} {282} (\bibinfo {year}
  {2012})}\BibitemShut {NoStop}%
\bibitem [{\citenamefont {Krockenberger}\ \emph {et~al.}(2013)\citenamefont
  {Krockenberger}, \citenamefont {Irie}, \citenamefont {Matsumoto},
  \citenamefont {Yamagami}, \citenamefont {Mitsuhashi}, \citenamefont
  {Tsukada}, \citenamefont {Naito},\ and\ \citenamefont
  {Yamamoto}}]{Krockenberger_13}%
  \BibitemOpen
  \bibfield  {author} {\bibinfo {author} {\bibfnamefont {Y.}~\bibnamefont
  {Krockenberger}}, \bibinfo {author} {\bibfnamefont {H.}~\bibnamefont {Irie}},
  \bibinfo {author} {\bibfnamefont {O.}~\bibnamefont {Matsumoto}}, \bibinfo
  {author} {\bibfnamefont {K.}~\bibnamefont {Yamagami}}, \bibinfo {author}
  {\bibfnamefont {M.}~\bibnamefont {Mitsuhashi}}, \bibinfo {author}
  {\bibfnamefont {A.}~\bibnamefont {Tsukada}}, \bibinfo {author} {\bibfnamefont
  {M.}~\bibnamefont {Naito}}, \ and\ \bibinfo {author} {\bibfnamefont
  {H.}~\bibnamefont {Yamamoto}},\ }\href@noop {} {\bibfield  {journal}
  {\bibinfo  {journal} {Scientific Reports}\ }\textbf {\bibinfo {volume} {3}},\
  \bibinfo {pages} {2235} (\bibinfo {year} {2013})}\BibitemShut {NoStop}%
\bibitem [{\citenamefont {Song}\ \emph {et~al.}(2012)\citenamefont {Song},
  \citenamefont {Park}, \citenamefont {Kim}, \citenamefont {Kim}, \citenamefont
  {Leem}, \citenamefont {Choi}, \citenamefont {Jung}, \citenamefont {Koh},
  \citenamefont {Han}, \citenamefont {Yoshida}, \citenamefont {Eisaki},
  \citenamefont {Lu}, \citenamefont {Shen},\ and\ \citenamefont
  {Kim}}]{Song_12}%
  \BibitemOpen
  \bibfield  {author} {\bibinfo {author} {\bibfnamefont {D.}~\bibnamefont
  {Song}}, \bibinfo {author} {\bibfnamefont {S.~R.}\ \bibnamefont {Park}},
  \bibinfo {author} {\bibfnamefont {C.}~\bibnamefont {Kim}}, \bibinfo {author}
  {\bibfnamefont {Y.}~\bibnamefont {Kim}}, \bibinfo {author} {\bibfnamefont
  {C.}~\bibnamefont {Leem}}, \bibinfo {author} {\bibfnamefont {S.}~\bibnamefont
  {Choi}}, \bibinfo {author} {\bibfnamefont {W.}~\bibnamefont {Jung}}, \bibinfo
  {author} {\bibfnamefont {Y.}~\bibnamefont {Koh}}, \bibinfo {author}
  {\bibfnamefont {G.}~\bibnamefont {Han}}, \bibinfo {author} {\bibfnamefont
  {Y.}~\bibnamefont {Yoshida}}, \bibinfo {author} {\bibfnamefont
  {H.}~\bibnamefont {Eisaki}}, \bibinfo {author} {\bibfnamefont {D.~H.}\
  \bibnamefont {Lu}}, \bibinfo {author} {\bibfnamefont {Z.-X.}\ \bibnamefont
  {Shen}}, \ and\ \bibinfo {author} {\bibfnamefont {C.}~\bibnamefont {Kim}},\
  }\href@noop {} {\bibfield  {journal} {\bibinfo  {journal} {Phys. Rev. B}\
  }\textbf {\bibinfo {volume} {86}},\ \bibinfo {pages} {144520} (\bibinfo
  {year} {2012})}\BibitemShut {NoStop}%
\bibitem [{\citenamefont {Onose}\ \emph {et~al.}(2004)\citenamefont {Onose},
  \citenamefont {Taguchi}, \citenamefont {Ishizaka},\ and\ \citenamefont
  {Tokura}}]{Onose_04}%
  \BibitemOpen
  \bibfield  {author} {\bibinfo {author} {\bibfnamefont {Y.}~\bibnamefont
  {Onose}}, \bibinfo {author} {\bibfnamefont {Y.}~\bibnamefont {Taguchi}},
  \bibinfo {author} {\bibfnamefont {K.}~\bibnamefont {Ishizaka}}, \ and\
  \bibinfo {author} {\bibfnamefont {Y.}~\bibnamefont {Tokura}},\ }\href@noop {}
  {\bibfield  {journal} {\bibinfo  {journal} {Phys. Rev. B}\ }\textbf {\bibinfo
  {volume} {69}},\ \bibinfo {pages} {024504} (\bibinfo {year}
  {2004})}\BibitemShut {NoStop}%
\bibitem [{\citenamefont {Kim}\ and\ \citenamefont {Gaskell}(1993)}]{Kim_93}%
  \BibitemOpen
  \bibfield  {author} {\bibinfo {author} {\bibfnamefont {J.}~\bibnamefont
  {Kim}}\ and\ \bibinfo {author} {\bibfnamefont {D.}~\bibnamefont {Gaskell}},\
  }\href@noop {} {\bibfield  {journal} {\bibinfo  {journal} {Physica C:
  Superconductivity}\ }\textbf {\bibinfo {volume} {209}},\ \bibinfo {pages}
  {381} (\bibinfo {year} {1993})}\BibitemShut {NoStop}%
\bibitem [{\citenamefont {Barone}\ \emph {et~al.}(2009)\citenamefont {Barone},
  \citenamefont {Guarino}, \citenamefont {Nigro}, \citenamefont {Romano},\ and\
  \citenamefont {Pagano}}]{Barone_09}%
  \BibitemOpen
  \bibfield  {author} {\bibinfo {author} {\bibfnamefont {C.}~\bibnamefont
  {Barone}}, \bibinfo {author} {\bibfnamefont {A.}~\bibnamefont {Guarino}},
  \bibinfo {author} {\bibfnamefont {A.}~\bibnamefont {Nigro}}, \bibinfo
  {author} {\bibfnamefont {A.}~\bibnamefont {Romano}}, \ and\ \bibinfo {author}
  {\bibfnamefont {S.}~\bibnamefont {Pagano}},\ }\href@noop {} {\bibfield
  {journal} {\bibinfo  {journal} {Phys. Rev. B}\ }\textbf {\bibinfo {volume}
  {80}},\ \bibinfo {pages} {224405} (\bibinfo {year} {2009})}\BibitemShut
  {NoStop}%
\bibitem [{\citenamefont {Barone}\ \emph {et~al.}(2011)\citenamefont {Barone},
  \citenamefont {Pagano}, \citenamefont {Guarino}, \citenamefont {Nigro},\ and\
  \citenamefont {Vecchione}}]{Barone_11}%
  \BibitemOpen
  \bibfield  {author} {\bibinfo {author} {\bibfnamefont {C.}~\bibnamefont
  {Barone}}, \bibinfo {author} {\bibfnamefont {S.}~\bibnamefont {Pagano}},
  \bibinfo {author} {\bibfnamefont {A.}~\bibnamefont {Guarino}}, \bibinfo
  {author} {\bibfnamefont {A.}~\bibnamefont {Nigro}}, \ and\ \bibinfo {author}
  {\bibfnamefont {A.}~\bibnamefont {Vecchione}},\ }\href@noop {} {\bibfield
  {journal} {\bibinfo  {journal} {Superconductor Science and Technology}\
  }\textbf {\bibinfo {volume} {24}},\ \bibinfo {pages} {085003} (\bibinfo
  {year} {2011})}\BibitemShut {NoStop}%
\bibitem [{\citenamefont {Liu}\ \emph {et~al.}(1993)\citenamefont {Liu},
  \citenamefont {Whitaker}, \citenamefont {Uher}, \citenamefont {Peng},
  \citenamefont {Li},\ and\ \citenamefont {Greene}}]{Liu_93}%
  \BibitemOpen
  \bibfield  {author} {\bibinfo {author} {\bibfnamefont {Y.}~\bibnamefont
  {Liu}}, \bibinfo {author} {\bibfnamefont {J.~F.}\ \bibnamefont {Whitaker}},
  \bibinfo {author} {\bibfnamefont {C.}~\bibnamefont {Uher}}, \bibinfo {author}
  {\bibfnamefont {J.}~\bibnamefont {Peng}}, \bibinfo {author} {\bibfnamefont
  {Z.~Y.}\ \bibnamefont {Li}}, \ and\ \bibinfo {author} {\bibfnamefont {R.~L.}\
  \bibnamefont {Greene}},\ }\href@noop {} {\bibfield  {journal} {\bibinfo
  {journal} {Applied Physics Letters}\ }\textbf {\bibinfo {volume} {63}},\
  \bibinfo {pages} {979} (\bibinfo {year} {1993})}\BibitemShut {NoStop}%
\bibitem [{\citenamefont {Okamoto}\ \emph {et~al.}(2010)\citenamefont
  {Okamoto}, \citenamefont {Miyagoe}, \citenamefont {Kobayashi}, \citenamefont
  {Uemura}, \citenamefont {Nishioka}, \citenamefont {Matsuzaki}, \citenamefont
  {Sawa},\ and\ \citenamefont {Tokura}}]{Okamoto_10}%
  \BibitemOpen
  \bibfield  {author} {\bibinfo {author} {\bibfnamefont {H.}~\bibnamefont
  {Okamoto}}, \bibinfo {author} {\bibfnamefont {T.}~\bibnamefont {Miyagoe}},
  \bibinfo {author} {\bibfnamefont {K.}~\bibnamefont {Kobayashi}}, \bibinfo
  {author} {\bibfnamefont {H.}~\bibnamefont {Uemura}}, \bibinfo {author}
  {\bibfnamefont {H.}~\bibnamefont {Nishioka}}, \bibinfo {author}
  {\bibfnamefont {H.}~\bibnamefont {Matsuzaki}}, \bibinfo {author}
  {\bibfnamefont {A.}~\bibnamefont {Sawa}}, \ and\ \bibinfo {author}
  {\bibfnamefont {Y.}~\bibnamefont {Tokura}},\ }\href@noop {} {\bibfield
  {journal} {\bibinfo  {journal} {Phys. Rev. B}\ }\textbf {\bibinfo {volume}
  {82}},\ \bibinfo {pages} {060513} (\bibinfo {year} {2010})}\BibitemShut
  {NoStop}%
\bibitem [{\citenamefont {Bonavolont{\`a}}\ \emph {et~al.}(2013)\citenamefont
  {Bonavolont{\`a}}, \citenamefont {Parlato}, \citenamefont {Pepe},
  \citenamefont {de~Lisio}, \citenamefont {Valentino}, \citenamefont
  {Bellingeri}, \citenamefont {Pallecchi}, \citenamefont {Putti},\ and\
  \citenamefont {Ferdeghini}}]{Bonavolonta_13}%
  \BibitemOpen
  \bibfield  {author} {\bibinfo {author} {\bibfnamefont {C.}~\bibnamefont
  {Bonavolont{\`a}}}, \bibinfo {author} {\bibfnamefont {L.}~\bibnamefont
  {Parlato}}, \bibinfo {author} {\bibfnamefont {G.~P.}\ \bibnamefont {Pepe}},
  \bibinfo {author} {\bibfnamefont {C.}~\bibnamefont {de~Lisio}}, \bibinfo
  {author} {\bibfnamefont {M.}~\bibnamefont {Valentino}}, \bibinfo {author}
  {\bibfnamefont {E.}~\bibnamefont {Bellingeri}}, \bibinfo {author}
  {\bibfnamefont {I.}~\bibnamefont {Pallecchi}}, \bibinfo {author}
  {\bibfnamefont {M.}~\bibnamefont {Putti}}, \ and\ \bibinfo {author}
  {\bibfnamefont {C.}~\bibnamefont {Ferdeghini}},\ }\href@noop {} {\bibfield
  {journal} {\bibinfo  {journal} {Superconductor Science and Technology}\
  }\textbf {\bibinfo {volume} {26}},\ \bibinfo {pages} {075018} (\bibinfo
  {year} {2013})}\BibitemShut {NoStop}%
\bibitem [{\citenamefont {Bonavolont{\`a}}\ \emph {et~al.}(2014)\citenamefont
  {Bonavolont{\`a}}, \citenamefont {de~Lisio}, \citenamefont {Valentino},
  \citenamefont {Parlato}, \citenamefont {Pepe}, \citenamefont {Kurth},\ and\
  \citenamefont {Iida}}]{Bonavolonta_14}%
  \BibitemOpen
  \bibfield  {author} {\bibinfo {author} {\bibfnamefont {C.}~\bibnamefont
  {Bonavolont{\`a}}}, \bibinfo {author} {\bibfnamefont {C.}~\bibnamefont
  {de~Lisio}}, \bibinfo {author} {\bibfnamefont {M.}~\bibnamefont {Valentino}},
  \bibinfo {author} {\bibfnamefont {L.}~\bibnamefont {Parlato}}, \bibinfo
  {author} {\bibfnamefont {G.}~\bibnamefont {Pepe}}, \bibinfo {author}
  {\bibfnamefont {F.}~\bibnamefont {Kurth}}, \ and\ \bibinfo {author}
  {\bibfnamefont {K.}~\bibnamefont {Iida}},\ }\href@noop {} {\bibfield
  {journal} {\bibinfo  {journal} {Physica C}\ }\textbf {\bibinfo {volume}
  {503}},\ \bibinfo {pages} {132} (\bibinfo {year} {2014})}\BibitemShut
  {NoStop}%
\bibitem [{\citenamefont {Cilento}\ \emph {et~al.}(2014)\citenamefont
  {Cilento}, \citenamefont {Dal~Conte}, \citenamefont {Coslovich},
  \citenamefont {Peli}, \citenamefont {Nembrini}, \citenamefont {Mor},
  \citenamefont {Banfi}, \citenamefont {Ferrini}, \citenamefont {Eisaki},
  \citenamefont {Chan}, \citenamefont {Dorow}, \citenamefont {Veit},
  \citenamefont {Greven}, \citenamefont {van~der Marel}, \citenamefont {Comin},
  \citenamefont {Damascelli}, \citenamefont {Rettig}, \citenamefont
  {Bovensiepen}, \citenamefont {Capone}, \citenamefont {Giannetti},\ and\
  \citenamefont {Parmigiani}}]{Cilento_14}%
  \BibitemOpen
  \bibfield  {author} {\bibinfo {author} {\bibfnamefont {F.}~\bibnamefont
  {Cilento}}, \bibinfo {author} {\bibfnamefont {S.}~\bibnamefont {Dal~Conte}},
  \bibinfo {author} {\bibfnamefont {G.}~\bibnamefont {Coslovich}}, \bibinfo
  {author} {\bibfnamefont {S.}~\bibnamefont {Peli}}, \bibinfo {author}
  {\bibfnamefont {N.}~\bibnamefont {Nembrini}}, \bibinfo {author}
  {\bibfnamefont {S.}~\bibnamefont {Mor}}, \bibinfo {author} {\bibfnamefont
  {F.}~\bibnamefont {Banfi}}, \bibinfo {author} {\bibfnamefont
  {G.}~\bibnamefont {Ferrini}}, \bibinfo {author} {\bibfnamefont
  {H.}~\bibnamefont {Eisaki}}, \bibinfo {author} {\bibfnamefont {M.~K.}\
  \bibnamefont {Chan}}, \bibinfo {author} {\bibfnamefont {C.~J.}\ \bibnamefont
  {Dorow}}, \bibinfo {author} {\bibfnamefont {M.~J.}\ \bibnamefont {Veit}},
  \bibinfo {author} {\bibfnamefont {M.}~\bibnamefont {Greven}}, \bibinfo
  {author} {\bibfnamefont {D.}~\bibnamefont {van~der Marel}}, \bibinfo {author}
  {\bibfnamefont {R.}~\bibnamefont {Comin}}, \bibinfo {author} {\bibfnamefont
  {A.}~\bibnamefont {Damascelli}}, \bibinfo {author} {\bibfnamefont
  {L.}~\bibnamefont {Rettig}}, \bibinfo {author} {\bibfnamefont
  {U.}~\bibnamefont {Bovensiepen}}, \bibinfo {author} {\bibfnamefont
  {M.}~\bibnamefont {Capone}}, \bibinfo {author} {\bibfnamefont
  {C.}~\bibnamefont {Giannetti}}, \ and\ \bibinfo {author} {\bibfnamefont
  {F.}~\bibnamefont {Parmigiani}},\ }\href@noop {} {\bibfield  {journal}
  {\bibinfo  {journal} {Nat. Commun.}\ }\textbf {\bibinfo {volume} {5}},\
  \bibinfo {pages} {4353} (\bibinfo {year} {2014})}\BibitemShut {NoStop}%
\bibitem [{\citenamefont {Novelli}\ \emph {et~al.}(2014)\citenamefont
  {Novelli}, \citenamefont {De~Filippis}, \citenamefont {Cataudella},
  \citenamefont {Esposito}, \citenamefont {Vergara}, \citenamefont {Cilento},
  \citenamefont {Sindici}, \citenamefont {Amaricci}, \citenamefont {Giannetti},
  \citenamefont {Prabhakaran}, \citenamefont {Wall}, \citenamefont {Perucchi},
  \citenamefont {Dal~Conte}, \citenamefont {Cerullo}, \citenamefont {Capone},
  \citenamefont {Mishchenko}, \citenamefont {Gr{\"u}ninger}, \citenamefont
  {Nagaosa}, \citenamefont {Parmigiani},\ and\ \citenamefont
  {Fausti}}]{Novelli_14}%
  \BibitemOpen
  \bibfield  {author} {\bibinfo {author} {\bibfnamefont {F.}~\bibnamefont
  {Novelli}}, \bibinfo {author} {\bibfnamefont {G.}~\bibnamefont
  {De~Filippis}}, \bibinfo {author} {\bibfnamefont {V.}~\bibnamefont
  {Cataudella}}, \bibinfo {author} {\bibfnamefont {M.}~\bibnamefont
  {Esposito}}, \bibinfo {author} {\bibfnamefont {I.}~\bibnamefont {Vergara}},
  \bibinfo {author} {\bibfnamefont {F.}~\bibnamefont {Cilento}}, \bibinfo
  {author} {\bibfnamefont {E.}~\bibnamefont {Sindici}}, \bibinfo {author}
  {\bibfnamefont {A.}~\bibnamefont {Amaricci}}, \bibinfo {author}
  {\bibfnamefont {C.}~\bibnamefont {Giannetti}}, \bibinfo {author}
  {\bibfnamefont {D.}~\bibnamefont {Prabhakaran}}, \bibinfo {author}
  {\bibfnamefont {S.}~\bibnamefont {Wall}}, \bibinfo {author} {\bibfnamefont
  {A.}~\bibnamefont {Perucchi}}, \bibinfo {author} {\bibfnamefont
  {S.}~\bibnamefont {Dal~Conte}}, \bibinfo {author} {\bibfnamefont
  {G.}~\bibnamefont {Cerullo}}, \bibinfo {author} {\bibfnamefont
  {M.}~\bibnamefont {Capone}}, \bibinfo {author} {\bibfnamefont
  {A.}~\bibnamefont {Mishchenko}}, \bibinfo {author} {\bibfnamefont
  {M.}~\bibnamefont {Gr{\"u}ninger}}, \bibinfo {author} {\bibfnamefont
  {N.}~\bibnamefont {Nagaosa}}, \bibinfo {author} {\bibfnamefont
  {F.}~\bibnamefont {Parmigiani}}, \ and\ \bibinfo {author} {\bibfnamefont
  {D.}~\bibnamefont {Fausti}},\ }\href@noop {} {\bibfield  {journal} {\bibinfo
  {journal} {Nat. Commun.}\ }\textbf {\bibinfo {volume} {5}},\ \bibinfo {pages}
  {5112} (\bibinfo {year} {2014})}\BibitemShut {NoStop}%
\bibitem [{\citenamefont {Li}\ \emph {et~al.}(2015)\citenamefont {Li},
  \citenamefont {Zhang}, \citenamefont {Wang}, \citenamefont {Chakhalian},\
  and\ \citenamefont {Xiao}}]{Li_15}%
  \BibitemOpen
  \bibfield  {author} {\bibinfo {author} {\bibfnamefont {W.}~\bibnamefont
  {Li}}, \bibinfo {author} {\bibfnamefont {C.}~\bibnamefont {Zhang}}, \bibinfo
  {author} {\bibfnamefont {X.}~\bibnamefont {Wang}}, \bibinfo {author}
  {\bibfnamefont {J.}~\bibnamefont {Chakhalian}}, \ and\ \bibinfo {author}
  {\bibfnamefont {M.}~\bibnamefont {Xiao}},\ }\href@noop {} {\bibfield
  {journal} {\bibinfo  {journal} {Journal of Magnetism and Magnetic Materials}\
  }\textbf {\bibinfo {volume} {376}},\ \bibinfo {pages} {29} (\bibinfo {year}
  {2015})}\BibitemShut {NoStop}%
\bibitem [{\citenamefont {Dal~Conte}\ \emph {et~al.}(2015)\citenamefont
  {Dal~Conte}, \citenamefont {Vidmar}, \citenamefont {Golez}, \citenamefont
  {Mierzejewski}, \citenamefont {Soavi}, \citenamefont {Peli}, \citenamefont
  {Banfi}, \citenamefont {Ferrini}, \citenamefont {Comin}, \citenamefont
  {Ludbrook}, \citenamefont {Chauviere}, \citenamefont {Zhigadlo},
  \citenamefont {Eisaki}, \citenamefont {Greven}, \citenamefont {Lupi},
  \citenamefont {Damascelli}, \citenamefont {Brida}, \citenamefont {Capone},
  \citenamefont {Bonca}, \citenamefont {Cerullo},\ and\ \citenamefont
  {Giannetti}}]{Dal-Conte_15}%
  \BibitemOpen
  \bibfield  {author} {\bibinfo {author} {\bibfnamefont {S.}~\bibnamefont
  {Dal~Conte}}, \bibinfo {author} {\bibfnamefont {L.}~\bibnamefont {Vidmar}},
  \bibinfo {author} {\bibfnamefont {D.}~\bibnamefont {Golez}}, \bibinfo
  {author} {\bibfnamefont {M.}~\bibnamefont {Mierzejewski}}, \bibinfo {author}
  {\bibfnamefont {G.}~\bibnamefont {Soavi}}, \bibinfo {author} {\bibfnamefont
  {S.}~\bibnamefont {Peli}}, \bibinfo {author} {\bibfnamefont {F.}~\bibnamefont
  {Banfi}}, \bibinfo {author} {\bibfnamefont {G.}~\bibnamefont {Ferrini}},
  \bibinfo {author} {\bibfnamefont {R.}~\bibnamefont {Comin}}, \bibinfo
  {author} {\bibfnamefont {B.~M.}\ \bibnamefont {Ludbrook}}, \bibinfo {author}
  {\bibfnamefont {L.}~\bibnamefont {Chauviere}}, \bibinfo {author}
  {\bibfnamefont {N.~D.}\ \bibnamefont {Zhigadlo}}, \bibinfo {author}
  {\bibfnamefont {H.}~\bibnamefont {Eisaki}}, \bibinfo {author} {\bibfnamefont
  {M.}~\bibnamefont {Greven}}, \bibinfo {author} {\bibfnamefont
  {S.}~\bibnamefont {Lupi}}, \bibinfo {author} {\bibfnamefont {A.}~\bibnamefont
  {Damascelli}}, \bibinfo {author} {\bibfnamefont {D.}~\bibnamefont {Brida}},
  \bibinfo {author} {\bibfnamefont {M.}~\bibnamefont {Capone}}, \bibinfo
  {author} {\bibfnamefont {J.}~\bibnamefont {Bonca}}, \bibinfo {author}
  {\bibfnamefont {G.}~\bibnamefont {Cerullo}}, \ and\ \bibinfo {author}
  {\bibfnamefont {C.}~\bibnamefont {Giannetti}},\ }\href@noop {} {\bibfield
  {journal} {\bibinfo  {journal} {Nat. Phys.}\ }\textbf {\bibinfo {volume}
  {11}},\ \bibinfo {pages} {421} (\bibinfo {year} {2015})}\BibitemShut
  {NoStop}%
\bibitem [{\citenamefont {Madan}\ \emph {et~al.}(2015)\citenamefont {Madan},
  \citenamefont {Kurosawa}, \citenamefont {Toda}, \citenamefont {Oda},
  \citenamefont {Mertelj},\ and\ \citenamefont {Mihailovic}}]{Madan_15}%
  \BibitemOpen
  \bibfield  {author} {\bibinfo {author} {\bibfnamefont {I.}~\bibnamefont
  {Madan}}, \bibinfo {author} {\bibfnamefont {T.}~\bibnamefont {Kurosawa}},
  \bibinfo {author} {\bibfnamefont {Y.}~\bibnamefont {Toda}}, \bibinfo {author}
  {\bibfnamefont {M.}~\bibnamefont {Oda}}, \bibinfo {author} {\bibfnamefont
  {T.}~\bibnamefont {Mertelj}}, \ and\ \bibinfo {author} {\bibfnamefont
  {D.}~\bibnamefont {Mihailovic}},\ }\href@noop {} {\bibfield  {journal}
  {\bibinfo  {journal} {Nat Commun}\ }\textbf {\bibinfo {volume} {6}} (\bibinfo
  {year} {2015})}\BibitemShut {NoStop}%
\bibitem [{\citenamefont {Hinton}\ \emph {et~al.}(2016)\citenamefont {Hinton},
  \citenamefont {Thewalt}, \citenamefont {Alpichshev}, \citenamefont {Mahmood},
  \citenamefont {Koralek}, \citenamefont {Chan}, \citenamefont {Veit},
  \citenamefont {Dorow}, \citenamefont {Barisic}, \citenamefont {Kemper},
  \citenamefont {Bonn}, \citenamefont {Hardy}, \citenamefont {Liang},
  \citenamefont {Gedik}, \citenamefont {Greven}, \citenamefont {Lanzara},\ and\
  \citenamefont {Orenstein}}]{Hinton_16}%
  \BibitemOpen
  \bibfield  {author} {\bibinfo {author} {\bibfnamefont {J.}~\bibnamefont
  {Hinton}}, \bibinfo {author} {\bibfnamefont {E.}~\bibnamefont {Thewalt}},
  \bibinfo {author} {\bibfnamefont {Z.}~\bibnamefont {Alpichshev}}, \bibinfo
  {author} {\bibfnamefont {F.}~\bibnamefont {Mahmood}}, \bibinfo {author}
  {\bibfnamefont {J.}~\bibnamefont {Koralek}}, \bibinfo {author} {\bibfnamefont
  {M.}~\bibnamefont {Chan}}, \bibinfo {author} {\bibfnamefont {M.}~\bibnamefont
  {Veit}}, \bibinfo {author} {\bibfnamefont {C.}~\bibnamefont {Dorow}},
  \bibinfo {author} {\bibfnamefont {N.}~\bibnamefont {Barisic}}, \bibinfo
  {author} {\bibfnamefont {A.}~\bibnamefont {Kemper}}, \bibinfo {author}
  {\bibfnamefont {D.}~\bibnamefont {Bonn}}, \bibinfo {author} {\bibfnamefont
  {W.}~\bibnamefont {Hardy}}, \bibinfo {author} {\bibfnamefont
  {R.}~\bibnamefont {Liang}}, \bibinfo {author} {\bibfnamefont
  {N.}~\bibnamefont {Gedik}}, \bibinfo {author} {\bibfnamefont
  {M.}~\bibnamefont {Greven}}, \bibinfo {author} {\bibfnamefont
  {A.}~\bibnamefont {Lanzara}}, \ and\ \bibinfo {author} {\bibfnamefont
  {J.}~\bibnamefont {Orenstein}},\ }\href@noop {} {\enquote {\bibinfo {title}
  {The rate of quasiparticle recombination probes the onset of coherence in
  cuprate superconductors},}\ } (\bibinfo {year} {2016}),\ \bibinfo {note}
  {arXiv:1601.05224}\BibitemShut {NoStop}%
\bibitem [{\citenamefont {Vishik}\ \emph {et~al.}(2016)\citenamefont {Vishik},
  \citenamefont {Mahmood}, \citenamefont {Alpichshev}, \citenamefont {Higgins},
  \citenamefont {Greene},\ and\ \citenamefont {Gedik}}]{Vishik_16}%
  \BibitemOpen
  \bibfield  {author} {\bibinfo {author} {\bibfnamefont {I.~M.}\ \bibnamefont
  {Vishik}}, \bibinfo {author} {\bibfnamefont {F.}~\bibnamefont {Mahmood}},
  \bibinfo {author} {\bibfnamefont {Z.}~\bibnamefont {Alpichshev}}, \bibinfo
  {author} {\bibfnamefont {J.}~\bibnamefont {Higgins}}, \bibinfo {author}
  {\bibfnamefont {R.~L.}\ \bibnamefont {Greene}}, \ and\ \bibinfo {author}
  {\bibfnamefont {N.}~\bibnamefont {Gedik}},\ }\href@noop {} {\enquote
  {\bibinfo {title} {Dynamics of quasiparticles and antiferromagnetic
  correlations in electron-doped cuprate la$_{2-x}$ce$_x$cuo$_{4\pmδ}$},}\ }
  (\bibinfo {year} {2016}),\ \bibinfo {note} {arXiv:1601.06694}\BibitemShut
  {NoStop}%
\bibitem [{\citenamefont {Giannetti}\ \emph {et~al.}(2016)\citenamefont
  {Giannetti}, \citenamefont {Capone}, \citenamefont {Fausti}, \citenamefont
  {Fabrizio}, \citenamefont {Parmigiani},\ and\ \citenamefont
  {Mihailovic}}]{Giannetti_16}%
  \BibitemOpen
  \bibfield  {author} {\bibinfo {author} {\bibfnamefont {C.}~\bibnamefont
  {Giannetti}}, \bibinfo {author} {\bibfnamefont {M.}~\bibnamefont {Capone}},
  \bibinfo {author} {\bibfnamefont {D.}~\bibnamefont {Fausti}}, \bibinfo
  {author} {\bibfnamefont {M.}~\bibnamefont {Fabrizio}}, \bibinfo {author}
  {\bibfnamefont {F.}~\bibnamefont {Parmigiani}}, \ and\ \bibinfo {author}
  {\bibfnamefont {D.}~\bibnamefont {Mihailovic}},\ }\href@noop {} {\enquote
  {\bibinfo {title} {Ultrafast optical spectroscopy of strongly correlated
  materials and high-temperature superconductors: a non-equilibrium
  approach},}\ } (\bibinfo {year} {2016}),\ \bibinfo {note}
  {arXiv:1601.07204}\BibitemShut {NoStop}%
\bibitem [{\citenamefont {Hinton}\ \emph {et~al.}(2013)\citenamefont {Hinton},
  \citenamefont {Koralek}, \citenamefont {Yu}, \citenamefont {Motoyama},
  \citenamefont {Lu}, \citenamefont {Vishwanath}, \citenamefont {Greven},\ and\
  \citenamefont {Orenstein}}]{Hinton_13}%
  \BibitemOpen
  \bibfield  {author} {\bibinfo {author} {\bibfnamefont {J.~P.}\ \bibnamefont
  {Hinton}}, \bibinfo {author} {\bibfnamefont {J.~D.}\ \bibnamefont {Koralek}},
  \bibinfo {author} {\bibfnamefont {G.}~\bibnamefont {Yu}}, \bibinfo {author}
  {\bibfnamefont {E.~M.}\ \bibnamefont {Motoyama}}, \bibinfo {author}
  {\bibfnamefont {Y.~M.}\ \bibnamefont {Lu}}, \bibinfo {author} {\bibfnamefont
  {A.}~\bibnamefont {Vishwanath}}, \bibinfo {author} {\bibfnamefont
  {M.}~\bibnamefont {Greven}}, \ and\ \bibinfo {author} {\bibfnamefont
  {J.}~\bibnamefont {Orenstein}},\ }\href@noop {} {\bibfield  {journal}
  {\bibinfo  {journal} {Phys. Rev. Lett.}\ }\textbf {\bibinfo {volume} {110}},\
  \bibinfo {pages} {217002} (\bibinfo {year} {2013})}\BibitemShut {NoStop}%
\bibitem [{\citenamefont {Merlin}(1997)}]{Merlin_97}%
  \BibitemOpen
  \bibfield  {author} {\bibinfo {author} {\bibfnamefont {R.}~\bibnamefont
  {Merlin}},\ }\href@noop {} {\bibfield  {journal} {\bibinfo  {journal} {Solid
  State Communications}\ }\textbf {\bibinfo {volume} {102}},\ \bibinfo {pages}
  {207} (\bibinfo {year} {1997})}\BibitemShut {NoStop}%
\bibitem [{\citenamefont {Yan}\ \emph {et~al.}(1985)\citenamefont {Yan},
  \citenamefont {Gamble},\ and\ \citenamefont {Nelson}}]{Yan_85}%
  \BibitemOpen
  \bibfield  {author} {\bibinfo {author} {\bibfnamefont {Y.}~\bibnamefont
  {Yan}}, \bibinfo {author} {\bibfnamefont {E.~B.}\ \bibnamefont {Gamble}}, \
  and\ \bibinfo {author} {\bibfnamefont {K.~A.}\ \bibnamefont {Nelson}},\
  }\href@noop {} {\bibfield  {journal} {\bibinfo  {journal} {The Journal of
  Chemical Physics}\ }\textbf {\bibinfo {volume} {83}},\ \bibinfo {pages}
  {5391} (\bibinfo {year} {1985})}\BibitemShut {NoStop}%
\bibitem [{\citenamefont {Brorson}\ \emph {et~al.}(1990)\citenamefont
  {Brorson}, \citenamefont {Kazeroonian}, \citenamefont {Moodera},
  \citenamefont {Face}, \citenamefont {Cheng}, \citenamefont {Ippen},
  \citenamefont {Dresselhaus},\ and\ \citenamefont {Dresselhaus}}]{Brorson_90}%
  \BibitemOpen
  \bibfield  {author} {\bibinfo {author} {\bibfnamefont {S.~D.}\ \bibnamefont
  {Brorson}}, \bibinfo {author} {\bibfnamefont {A.}~\bibnamefont
  {Kazeroonian}}, \bibinfo {author} {\bibfnamefont {J.~S.}\ \bibnamefont
  {Moodera}}, \bibinfo {author} {\bibfnamefont {D.~W.}\ \bibnamefont {Face}},
  \bibinfo {author} {\bibfnamefont {T.~K.}\ \bibnamefont {Cheng}}, \bibinfo
  {author} {\bibfnamefont {E.~P.}\ \bibnamefont {Ippen}}, \bibinfo {author}
  {\bibfnamefont {M.~S.}\ \bibnamefont {Dresselhaus}}, \ and\ \bibinfo {author}
  {\bibfnamefont {G.}~\bibnamefont {Dresselhaus}},\ }\href@noop {} {\bibfield
  {journal} {\bibinfo  {journal} {Phys. Rev. Lett.}\ }\textbf {\bibinfo
  {volume} {64}},\ \bibinfo {pages} {2172} (\bibinfo {year}
  {1990})}\BibitemShut {NoStop}%
\bibitem [{\citenamefont {Lee}\ and\ \citenamefont
  {Ramakrishnan}(1985)}]{Lee_85}%
  \BibitemOpen
  \bibfield  {author} {\bibinfo {author} {\bibfnamefont {P.~A.}\ \bibnamefont
  {Lee}}\ and\ \bibinfo {author} {\bibfnamefont {T.~V.}\ \bibnamefont
  {Ramakrishnan}},\ }\href@noop {} {\bibfield  {journal} {\bibinfo  {journal}
  {Rev. Mod. Phys.}\ }\textbf {\bibinfo {volume} {57}},\ \bibinfo {pages} {287}
  (\bibinfo {year} {1985})}\BibitemShut {NoStop}%
\end{thebibliography}

%

\end{document}